\documentclass[]{aastex63}

\submitjournal{PSJ}

\shorttitle{Enceladus' Ocean}
\shortauthors{Zeng \& Jansen}

\graphicspath{{./}{figures/}}

\begin{document}

\title{Ocean Circulation on Enceladus With a High Versus Low Salinity Ocean}

\correspondingauthor{Yaoxuan Zeng}
\email{yxzeng@pku.edu.cn}

\author[0000-0002-2624-8579]{Yaoxuan Zeng}
\affiliation{Department of Atmospheric and Oceanic Sciences, School of Physics, Peking University, Beijing 100871, China.}

\author[0000-0002-6479-8651]{Malte F. Jansen}
\affiliation{Department of the Geophysical Sciences, University of Chicago, Chicago, IL 60637, USA.}

\begin{abstract}
Previous studies that have considered the ocean circulation on Enceladus have generally assumed the salinity to be Earth-like. However, according to observations and geochemical constraints, the salinity of Enceladus' ocean is likely to be lower, and importantly, it is probably low enough to reverse the sign of thermal expansivity. We investigate the ocean circulation and stratification of Enceladus' ocean using a combination of theoretical arguments and simulations using the MITgcm. We find that, if the salinity is high, the whole ocean is unstratified, and convection dominates the entire ocean. However, if the salinity is low enough, there exists a stratified layer in the upper ocean, whose thickness depends on the magnitude of the turbulent vertical diffusivity, which remains poorly constrained. Such a layer can suppress the vertical flux of heat and tracers, thereby affecting the heat flux to the ice shell and leading to a vertical tracer mixing time scale across the stratified layer of at least hundreds of years. This time scale is inconsistent with a previous estimate of vertical ocean mixing of several years, based on the size of detected silica nanoparticles in the plumes, leading us to conclude that either the salinity of Enceladus' ocean is higher than previously suggested or the interpretation of silica nanoparticle observations has to be reconsidered.
\end{abstract}

\keywords{Enceladus | Ocean circulation | Stratified layer | Ocean heat transport | Tracer mixing time scale}

\section{Introduction} \label{sec:intro}

Strong evidence suggests that Enceladus maintains a global ocean \citep[e.g.,][]{postberg2011salt, patthoff2011fracture, thomas2016enceladus}, and the possible existence of liquid water in contact with a rocky interior makes it a hot target in the search for life in the solar system. Material from Enceladus' ocean is continuously ejected into space, where some of it forms Saturn's E-ring \citep{schmidt2008slow, kempf2008ring}, which makes it the only extraterrestrial ocean that is known to be so accessible for sampling. Understanding the ocean on Enceladus can also assist in understanding oceans on other icy moons in and outside the solar system, and help us better predict their habitability.

The ocean on Enceladus is estimated to be about 40~km in depth on average, and covered by a global ice shell \citep{thomas2016enceladus}. The ice shell is about 20~km deep on average, with the thickest part at the equator estimated to be more than 30~km and the thinnest part at the south pole less than 10~km \citep{beuthe2016enceladus, vcadek2019long, hemingway2019enceladus}. To maintain such a global ocean under the ice shell, as well as to explain the heat loss rate of around 10~GW in the south polar region \citep{spencer2006cassini, howett2011high, spencer2013enceladus}, there must be an energy source inside Enceladus. 

The energy source is likely to be associated primarily with tidal dissipation. In general, tidal dissipation is expected to occur in the ice shell, in the ocean, and in the solid core. However, tidal heating in the ocean is believed to be negligible compared to that in the ice shell and inner solid core \citep{chen2011obliquity, tyler2011tidal, beuthe2016crustal, hay2017numerically}. Libration can also generate heat in the ocean and may result in a total heating of up to O(0.1 GW) \citep[e.g.,][]{wilson2018can, rekier2019internal, soderlund2020ice}, which, however, is still smaller than the estimated tidal dissipation rate in the ice shell and the solid core. The total tidal dissipation rate in the ice shell (not including dissipation within liquid-water conduits) has recently been suggested to be on the order of 1~GW, with the maximum at the south pole where the ice shell is thinnest \citep{souvcek2019tidal, beuthe2019enceladus}. Tidal dissipation in the ice shell has also been suggested as an explanation for the north-south asymmetry \citep{kang2020spontaneous} and equator-to-pole variations \citep{kang2020differing} in the ice shell thickness, as well as for the sustained plumes associated with the south-polar tiger stripes \citep{kite2016sustained}. Continuous high-temperature hydrothermal activity suggests vigorous tidal heating in the solid core \citep{sekine2015high, hsu2015ongoing}. The tidal dissipation rate in the core is likely to reach O(10~GW), and is believed to be strongest at the pole and weakest at the substellar and anti-substellar point at the equator \citep{choblet2017powering}. Tidal energy dissipation in the solid core will generate heat, which must be transported outwards to the ice shell by the ocean circulation. Although the fraction of tidal heating coming from the solid core and the ice shell remains highly uncertain, some heating from the bottom solid core is expected, which fundamentally shapes the ocean circulation on Enceladus.

Previous studies have looked into possible scenarios for the ocean circulation on Enceladus and other icy moons, with both ocean-only models and ice-ocean coupled models \citep[e.g.,][]{soderlund2014ocean, travis2015keeping, soderlund2019ocean, amit2020cooling, ashkenazy2020europa, kang2020differing, kang2021does}. One characteristic feature of the ocean circulation on Enceladus may be hydrothermal convection columns, which are expected to be aligned parallel to the rotation axis, and could extend from the sea floor to the ice-ocean interface \citep{goodman2004hydrothermal, goodman2012numerical, soderlund2019ocean, ashkenazy2020europa, kang2020differing}. For another icy moon, Europa, hydrothermal convection plumes, quasi-3D turbulence, and baroclinic eddies have been suggested, based on different estimated parameter regimes, characterizing the strengths of the forcing that drives convection \citep{goodman2004hydrothermal, goodman2012numerical, soderlund2014ocean}, and depending on whether salinity changes from freezing and melting are taken into consideration \citep{ashkenazy2020europa}. Noting the substantial discrepancies among the simulations, the large uncertainty in external parameters, and the inherent computational challenges in modelling the global ocean in a realistic parameter regime, the ocean circulation regime on Enceladus remains uncertain.

\begin{figure}[b]
\centering
\includegraphics[width=0.9\linewidth]{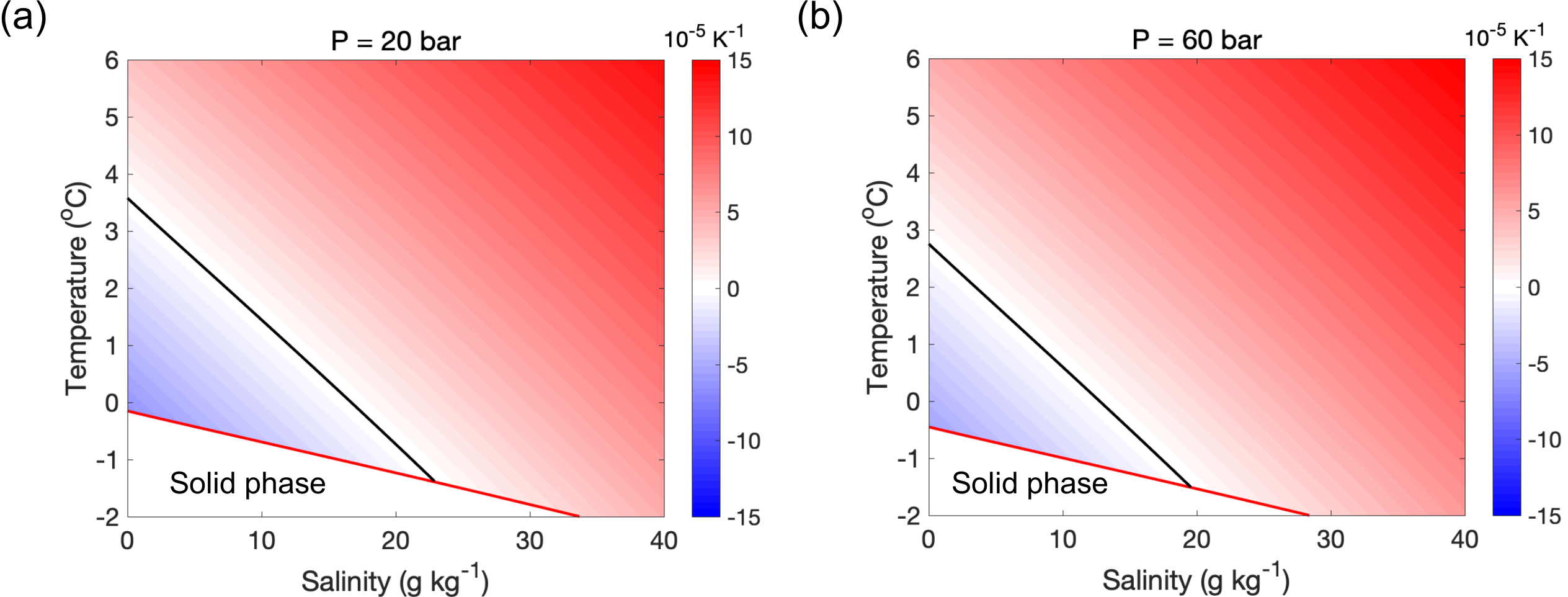}
\caption{Thermal expansivity $\alpha = -(1/\rho)(\partial \rho / \partial T)$ of sea water as a function of salinity and temperature at (a) 20~bar and (b) 60~bar pressure, representative of the top and bottom of Enceladus' ocean, respectively. The black line indicates the zero contour. The red line indicates the freezing point $T_f$ of sea water as a function of salinity. The area below the black line is where thermal expansivity of sea water is negative. The salinity at the intersection of the black and red lines is the maximum salinity for which negative $\alpha$ can exist under the given pressure. The calculation of the density is based on \citet{jackett1995minimal}, and the calculation of the freezing point is based on \citet{fofonoff1983algorithms}.}
\label{fig_salt}
\end{figure}

All previous simulations of Enceladus' ocean assumed the salinity to be roughly similar to Earth's ocean and thus high enough that the thermal expansivity ($\alpha = -(1/\rho)(\partial \rho / \partial T$)) is always positive. However, the ocean on Enceladus is likely to be fresher than Earth's ocean. Although some studies have suggested the salinity to be higher than 20~g~kg$^{-1}$ \citep[e.g.,][]{ingersoll2016controlled} or around 20~g~kg$^{-1}$ \citep{kang2021does}, based on dynamical considerations, geochemical evidence and modelling suggest that a salinity of around 20~g~kg$^{-1}$, dominated by NaCl, is an upper bound \citep[e.g.,][]{postberg2009sodium, hsu2015ongoing, glein2018geochemistry}. If the salinity is less than 20~g~kg$^{-1}$, the thermal expansivity of sea water is negative near the freezing point (Figure~\ref{fig_salt}). In this case, a stably stratified layer may exist in the upper ocean on Enceladus, similar to a scenario that has been proposed for Europa by \citet{melosh2004temperature}. To the best of our knowledge, no General Circulation Model (GCM) simulations have thus far been carried out to confirm the existence and effect of such a stably stratified layer.

\setcounter{footnote}{2}

Here we investigate the role of salinity on the dynamics of Enceladus' ocean, focusing in particular on the ocean's role in transporting heat and tracers from the sea floor to the ice shell. Ocean mixing can play a key role in transporting heat and constituents from the ocean-rock interface to the ice shell, thus affecting ice melting as well as properties observable in plumes. According to the detected size of silica nanoparticles and the growth rate expected from Ostwald ripening, the vertical mixing time scale has been estimated to be at most several years\footnote{Throughout this manuscript we will use years (and similarly months and days) as referring to Earth-years.} \citep{hsu2015ongoing}. One goal of this study is to analyze whether this vertical mixing time scale is consistent with the expected ocean circulation and transport on Enceladus. We present theoretical arguments and carry out global ocean simulations using the 3D general circulation model MITgcm. Section~\ref{sec:theory} describes theoretical predictions for the ocean circulation on Enceladus. Section~\ref{sec:simulations} shows results from numerical simulations. Section~\ref{sec:conclusion} provides concluding remarks.

\section{Theoretical Analysis} \label{sec:theory}

\subsection{The Thermal Expansion Coefficient and Implications for Ocean Stratification} \label{subsec:stratification_theory}

\begin{figure}[b]
\centering
\includegraphics[width=0.9\linewidth]{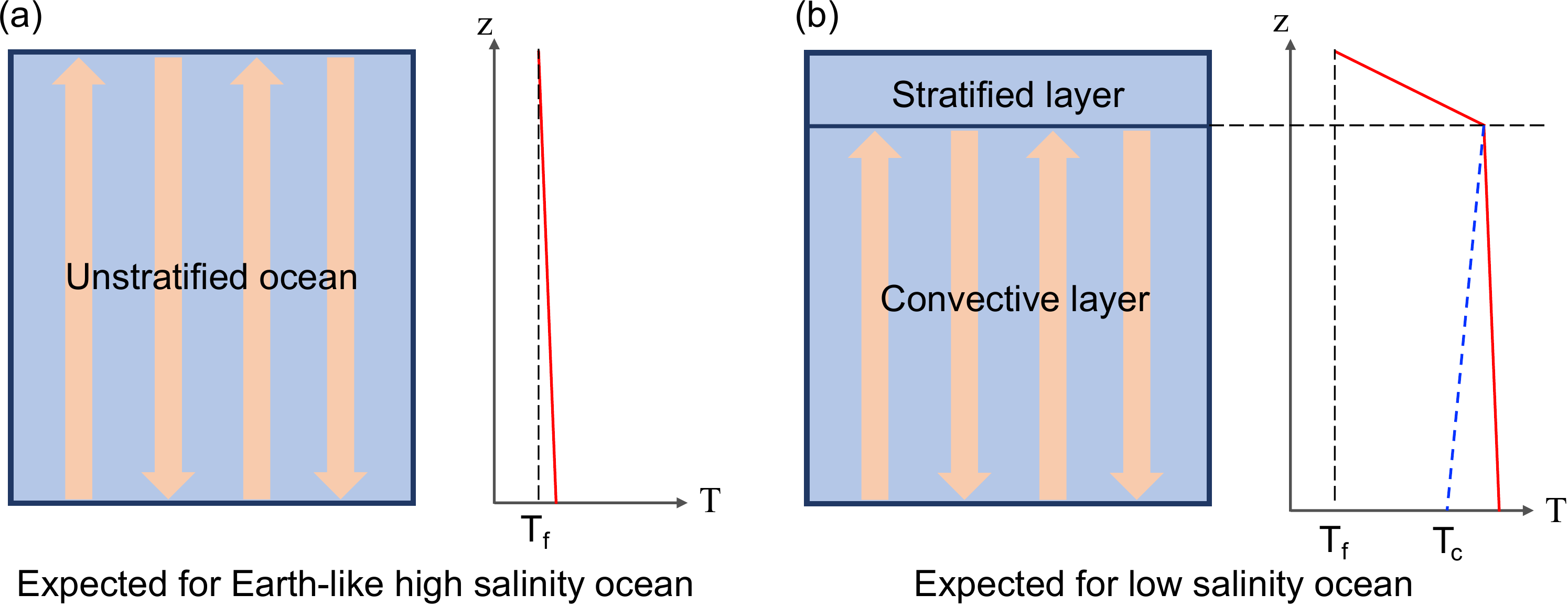}
\caption{Schematic for the expected vertical structure of the ocean on Enceladus with different salinities. (a) and (b) show the vertical structure and temperature profile of high and low salinity oceans, respectively. Red lines indicate the temperature profile, and the blue dashed line in (b) indicates the critical temperature $T_c$, which decreases as pressure increases. The high salinity ocean is expected to be virtually unstratified, with convection throughout the ocean and a small negative vertical temperature gradient around the freezing point. In the low salinity ocean, we expect two layers: the upper stratified layer with linear vertical temperature profile, and the lower convective layer that has a weak vertical temperature gradient. The temperature at the interface between these two layers is at the critical temperature $T_c$, where the thermal expansion coefficient changes sign. (See Figure~\ref{fig_HeatModel} and associated discussions for the horizontal structure of the heat flux.)}
\label{fig_vertical}
\end{figure}

The equation of state for saltwater describes the density as a non-linear function of temperature ($T$), salinity ($S$), and pressure ($P$) \citep{jackett1995minimal}. For salinities and pressures found in Earth's ocean, the density increases with salinity and decreases with temperature. Assuming a heat flux from the bottom due to tidal energy dissipation in the rocky interior, convection may then be expected in the whole ocean. As a result, the ocean would likely be well-mixed with very small temperature gradients (Figure~\ref{fig_vertical}(a)), consistent with the simulation results of \citet{soderlund2019ocean}, \citet{ashkenazy2020europa}, and \citet{kang2020differing}.

However, for low enough temperature and salinity (and at the modest pressures expected on Enceladus), the thermal expansion coefficient, $\alpha$, is negative, such that density increases with temperature. Under such circumstance, a critical temperature, $T_c$, exists at which the density of water reaches a maximum at a given salinity and pressure (Figure~\ref{fig_salt}). In this regime, heating from below will not trigger convection until the temperature is heated above $T_c$. However, the temperature at the upper boundary will be kept at the freezing point, $T_f$, due to the surface ice shell. A stably stratified layer is thus expected to form in the upper ocean, where the thermal expansion coefficient is negative and the temperature decreases upwards from $T_c$ to $T_f$. This is similar to the stratified layer that has been hypothesized to exist in the ocean of Europa under the low salinity assumption by \citet{melosh2004temperature}, although we note that the lower gravity and hence lower pressures on Enceladus make the conditions more favorable for the existence of a stably stratified layer. Below the stratified layer, the temperature is above $T_c$ so that the thermal expansivity is positive. This layer is expected to be qualitatively similar to the high salinity scenario and we expect it to be characterized by convection. The temperature at the interface between these two layers should be near the critical temperature $T_c$. A schematic for the expected vertical stratification in a low salinity ocean is shown in Figure~\ref{fig_vertical}(b).

Since convection cannot occur in the stratified layer, the vertical heat flux is expected to be dominated by diffusion (driven by either molecular diffusion or small-scale turbulence). We can then estimate the depth of the stratified layer using the equation of heat diffusion, $Q=c_p \rho \kappa_{z,heat} \partial T/ \partial z$, where $Q$ is the vertical heat flux, $c_p$ is the heat capacity, $\rho$ is the density of the water, and $\kappa_{z,heat}$ is the vertical thermal diffusion coefficient. Given the temperature contrast across the stratified layer, $\Delta T = T_f - T_c$, we can estimate the depth of the stratified layer $H$ as

\begin{equation} \label{eq:H}
    H = \frac{c_p \rho \kappa_{z,heat} \left| \Delta T \right|}{Q}.
\end{equation}

If we assume the total bottom heating to be around 20~GW \citep{choblet2017powering}, the mean vertical heat flux is $Q\approx 0.03$~W~m$^{-2}$ at the top of the ocean. This value is consistent with the magnitude estimated in \citet{vcadek2019long} (i.e. tens of milliwatts per squaremeter). If we further assume the salinity to be 8.5~g~kg$^{-1}$ for Enceladus' ocean \citep{glein2018geochemistry}, we find $T_f\approx -0.6^\circ$C, $T_c\approx 1.4^\circ$C, so that $\left| \Delta T \right| \approx $2.0~K. If we use the magnitude of the molecular thermal diffusivity, around 10$^{-7}$ m$^2$ s$^{-1}$, the depth of the stratified layer is $H\approx 30$~m. In general, vertical mixing in the ocean can be intensified through turbulence, in which case the molecular diffusivity should be replaced by the turbulent diffusivity. If we take the magnitude of the turbulent diffusivity in Earth's ocean, around $10^{-5}$~m$^2$~s$^{-1}$ \citep{munk1998abyssal}, the depth of the stratified layer would be $H\approx 3$~km. The question of whether and by how much vertical mixing in the ocean of Enceladus is enhanced by turbulence is hence important and is discussed in the following section.

\subsection{Tidal Dissipation and Turbulent Mixing in a Stratified Ocean} \label{subsec:tidal}

Vertical mixing in a stably stratified ocean increases the potential energy of the water column as buoyancy is fluxed downward (mixing lighter water downwards and denser water upwards), and hence requires a source of energy. If we know the rate of turbulent kinetic energy dissipation in the ocean's interior, we can estimate the vertical turbulent diffusivity based on the energy required to mix a stably stratified ocean \citep[e.g.,][]{wunsch2004vertical, yang2017persistence}:

\begin{equation} \label{eq:kappa1}
    \kappa_z = \frac{\Gamma \varepsilon}{\rho N^2}.
\end{equation}

Here $\kappa_z$ is the vertical turbulent diffusivity, $\varepsilon$ is the turbulent kinetic energy dissipation per unit volume, $\Gamma\approx 20\%$ is the ``mixing efficiency'', i.e. the fraction of the kinetic energy dissipation that contributes to the generation of potential energy \citep{peltier2003mixing, wunsch2004vertical}, and $N$ is the Brunt-V\"ais\"al\"a frequency. Given that $N^2 = -(g/\rho_{\theta})(\partial \rho_{\theta} / \partial z) \approx \alpha g\Delta T/H$ where $\rho_{\theta}$ is potential density and $g$ is gravity, we have

\begin{equation}\label{eq:kappa2}
    \kappa_z = \frac{\Gamma \varepsilon H}{\rho g \alpha \Delta T} = \frac{\Gamma  E}{\alpha \rho g \Delta T A},
\end{equation}

\noindent where $E$ is the total turbulent kinetic energy dissipation rate and $A$ is the horizontal area, and hence $\Gamma E/A$ is the energy used for vertical mixing per unit area, in units of W~m$^{-2}$.

On Enceladus, the turbulent kinetic energy that drives vertical mixing of the stably stratified water column can be derived from tidal disspation and libration. The tidal dissipation in the ocean on Enceladus is likely to be relatively weak, and is suggested to be 10$^1$-$10^4$~W, based on different models \citep{chen2014tidal, matsuyama2018ocean, hay2019nonlinear}. Libration can result in significant dissipation up to O(0.1 GW) in the ocean \citep{wilson2018can}. The dissipation caused by libration is likely to be concentrated primarily in the surface Ekman boundary layer \citep[e.g.,][]{greenspan1968theory, liao2008viscous, cebron2012elliptical, lemasquerier2017libration}. If mixing is confined to a thin boundary layer under the ice shell, it is expected to have little effect on the stratified layer underneath. However, some energy may be transferred into internal waves which can break and contribute to mixing in the interior \citep[e.g.,][]{sutherland1994internal, sutherland1994turbulence, sutherland1995internal, werne1999stratified, wunsch2004vertical}. Whether the libration-driven dissipation is concentrated in the surface boundary layer, or can significantly contribute to interior mixing, remains uncertain. We here treat the value of 0.1 GW as an upper-bound limit for the interior turbulent energy dissipation that contributes to mixing in the stratified layer, although we consider a much smaller value to be more likely.

We therefore estimate that the total turbulent energy dissipation rate that contributes to mixing of the stratified layer may be anywhere in the range of $E\sim 10^1-10^8$~W, and thus the energy input to vertical mixing per unit area, $\Gamma E/A$, is around $3 \times 10^{-12}$ to $3 \times 10^{-5}$~W~m$^{-2}$. Following the estimate in Section~\ref{subsec:stratification_theory} for the conditions in the stratified layer of the low salinity ocean, we get $\Delta T \approx -2.0$~K and $\alpha \approx -4 \times 10^{-5}$~K$^{-1}$, where we assumed a pressure of 20 bar and a temperature near the freezing point. We can then estimate  $\kappa_z$ to be around $3 \times 10^{-10}$ to $3 \times 10^{-3}$~m$^2$~s$^{-1}$. For the upper limit ($3 \times 10^{-3}$~m$^2$~s$^{-1}$), the depth of the stratified layer would be expected to be around 800 km according to Equation~(\ref{eq:H}), which is deeper than the whole ocean. This indicates that if libration can drive strong vertical turbulent mixing in the interior, the whole ocean may become stably stratified, as seen in the low salinity simulations of \citet{kang2021does} where a vertical diffusivity of $5 \times 10^{-3}$~m$^2$~s$^{-1}$ is assumed. For the lower limit, the turbulent diffusivity is weaker than the molecular value (10$^{-7}$~m$^2$~s$^{-1}$ for thermal diffusion and 10$^{-9}$~m$^2$~s$^{-1}$ for tracer diffusion), indicating that turbulence would be unable to significantly enhance vertical mixing in the interior.

\subsection{Conceptual Models of Ocean Heat Transport} \label{subsec:HeatModel}

To illustrate the role of ocean dynamics for the transport of heat from the rocky core to the ice shell, we here discuss conceptual models of ocean heat transport based on possible directions of anisotropy in mixing brought about by rotation and gravity. Anisotropy brought about by gravity tends to align radially, while anisotropy due to rotation tends to align along the rotation axis (like Taylor columns). We therefore suggest four conceptual models in which heat is transported along/normal to the direction of rotation/gravity.

\begin{figure}[b]
\centering
\includegraphics[width=0.9\linewidth]{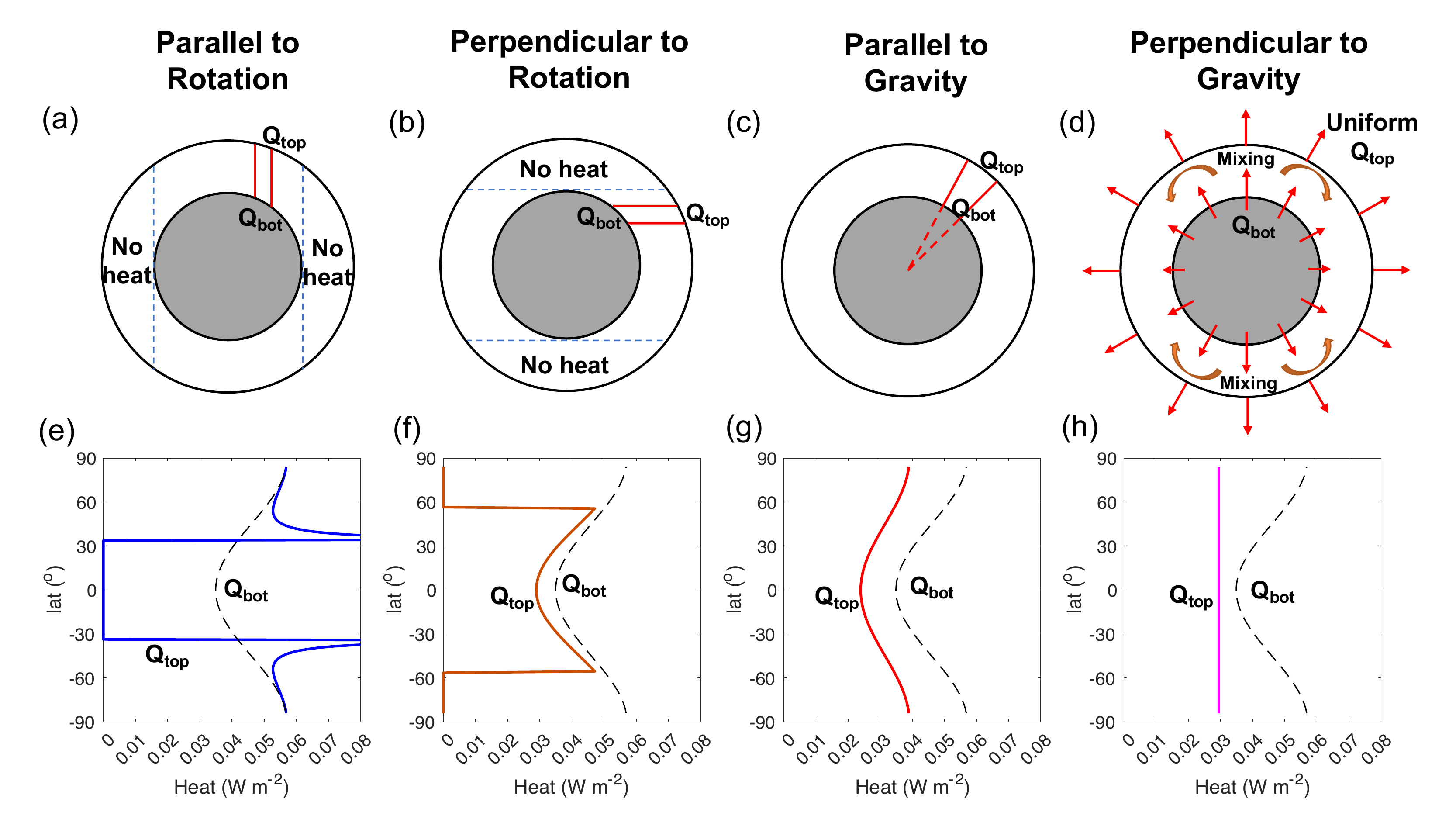}
\caption{Four conceptual models of ocean heat transport. (a)-(d) are schematics and (e)-(h) are the expected bottom and surface heat fluxes in the four models: mixing parallel to the direction of rotation, perpendicular to the direction of rotation, parallel to the direction of gravity, and perpendicular to the direction of gravity. In (e)-(h), the dashed lines indicate the prescribed bottom heat fluxes and the solid lines indicate the surface heat fluxes. For more details on the calculation of the surface heat fluxes, see APPENDIX \ref{si:heat}.}
\label{fig_HeatModel}
\end{figure}

In the first model ``Parallel to Rotation'', we assume heat is transported along the axis of rotation, representing slantwise convection (Figure~\ref{fig_HeatModel}(a)). On Enceladus, convection is likely to be parallel to the rotation axis--or more generally along surfaces of constant total angular momentum \citep{goodman2004hydrothermal, goodman2012numerical, soderlund2019ocean, kang2020differing}. In this limit, the heat flux inside the tangent cylinder that encircles the solid core (the blue dashed lines in Figure~\ref{fig_HeatModel}(a)), can be calculated by matching the bottom heat flux to the corresponding area at the ice-ocean interface. At the intersection of the tangent cylinder and the ocean surface, the heat flux to the ice-ocean interface diverges in this model, because heat from a finite bottom area is transported to an infinitesimal surface area. Outside the tangent cylinder (at low-latitudes), there is no heat flux at the surface of the ocean since slantwise convective columns, following the axis of rotation, cannot reach the surface there (Figure~\ref{fig_HeatModel}(e)).

In the second model ``Perpendicular to Rotation'', we consider the limit case where all heat is transported in columns perpendicular to the rotation axis (Figure~\ref{fig_HeatModel}(b)). This model represents the lateral heat transport by Taylor columns, which may be dominant on icy moons, especially at low latitudes \citep[e.g.,][]{cardin1994chaotic, christensen2002zonal, kang2020differing}. Opposite to the first model, there is no heat flux at high latitudes while the heat flux at low latitudes is calculated by matching the bottom heat flux to the corresponding area at the ice-ocean interface (Figure~\ref{fig_HeatModel}(f)). To the best of our knowledge, a vanishingly small heat flux to the ice shell at the equator or the pole has not been observed in any prior simulations of icy moon oceans, although many previous simulations do show differing heat fluxes at low- versus high-latitudes \citep[e.g.,][]{soderlund2019ocean, amit2020cooling, kang2020differing}, with strongly differing patterns resulting from different assumptions for various parameters. Specifically, the results of \citet{amit2020cooling} suggest that heat flux parallel to the axis of rotation may be dominant at relatively large Rossby numbers, while heat flux perpendicular to the axis of rotation may be expected to dominate at relatively small Rossby numbers. In general, we may not expect one single model/mechanism to fully explain the heat transport process, but, depending on parameter assumptions, some combination of these models may be applicable.

In the third model, ``Parallel to Gravity'', we assume that heat is transported to the surface radially. This model is expected to be relevant if heat transport is dominated either by radial convection, or by 3-dimensional turbulence or molecular diffusion, which generate an approximately isotropic diffusive transport. If the depth of the ocean is relatively small compared to the horizontal scale over which temperature varies, the heat flux in isotropic diffusion is expected to be dominated by the radial component. In this limit, the pattern of the heat flux at the top of the ocean will match the bottom heat flux, reduced by a constant factor to account for the differences between the area at the bottom and the surface (Figure~\ref{fig_HeatModel}(c)~\&~(g)).

In the fourth model ``Perpendicular to Gravity'', we assume that heat is well mixed horizontally (normal to the direction of gravity). If horizontal mixing is very efficient (e.g. due to the presence of geostrophic turbulence), any gradients in the deep ocean heating should be homogenized by ocean mixing and a relatively uniform surface heat flux is expected (Figure~\ref{fig_HeatModel}(d)). Under this circumstance, the surface heat flux should be equal to the average of the bottom heat flux (reduced by a constant factor) everywhere (Figure~\ref{fig_HeatModel}(h)). 

The heat flux to the surface ice shell is likely to play a role in determining the ice thickness distribution. With more heat transported to the surface at the pole than at the equator (the ``Parallel to Rotation'' and ``Parallel to Gravity'' models), ice is likely to be melting at the pole but forming at the equator. This heat flux pattern is therefore qualitatively consistent with the observed ice thickness distribution, and is also qualitatively similar to the heat flux in the low Ekman number simulation of \citet{soderlund2019ocean}, and the relatively large Rossby number simulations of \citet{amit2020cooling}. With strong horizontal ocean heat transport (the ``Perpendicular to Gravity'' model), the thickness of the ice shell may be expected to be uniform \citep{ashkenazy2018dynamics}. This heat flux pattern is qualitatively similar to that found in the simulations of Europa's ocean in \citet{ashkenazy2020europa}. The ``Perpendicular to Rotation'' and the ``Perpendicular to Gravity'' model cannot explain the observation that the ice shell is thicker at the pole than the equator on Enceladus \citep{vcadek2019long, hemingway2019enceladus}, and would require the ice thickness distribution to be shaped by tidal dissipation in the ice shell itself \citep[e.g.,][]{kang2020spontaneous, kang2020differing}.

\subsection{Vertical Tracer Mixing Time Scale} \label{subsec:tracer_theory}

In the high salinity regime (or the convective layer in the low salinity regime), geochemical tracers and small particles can be transported from the bottom to the surface of the ocean (or the top of the convective layer) through convective processes relatively effectively. The vertical velocity in a rotating plume can be estimated as $w\sim \sqrt{B/f}$, where $B = g \alpha Q/ \rho c_p$ is the buoyancy flux and $f\approx 1\times 10^{-4}$~s$^{-1}$ is the Coriolis parameter \citep{jones1993convection}. The convective mixing time scale is then estimated to be

\begin{equation}\label{eq:convection}
    \tau_{conv} \sim \frac{D}{w} = \left(\frac{D^2f}{B}\right)^{1/2} = \left(\frac{\rho c_p f D^2}{\alpha g Q}\right)^{1/2},
\end{equation}

\noindent where $D \approx 40$~km is estimated as the depth of the ocean. If we assume the heat flux $Q \approx 0.03$~W~m$^{-2}$ and thermal expansivity $\alpha$ to be around $10^{-6}-10^{-4}$~K$^{-1}$, we find $\tau_{conv}\approx 50-500$~years. For an ocean with Earth-like salinity (35 g~kg$^{-1}$), the thermal expansivity is relatively large and the lower limit (tens of years) is more applicable.

However, in the low salinity regime, there is no convection in the stratified layer, where vertical transport of tracers is achieved mainly by diffusion (either turbulent or molecular). The diffusive time scale in the stratified layer can be estimated by the vertical diffusion equation:

\begin{equation}\label{eq:diff}
    \frac{\partial C}{\partial t} = \frac{\partial}{\partial z}\left(\kappa_{z,tracer} \frac{\partial C}{\partial z}\right),
\end{equation}

\noindent where $C$ is the tracer concentration and $\kappa_{z,tracer}$ is the vertical diffusivity of the tracer. Through scaling analysis of Equation (\ref{eq:diff}), we find the diffusive mixing time scale

\begin{equation} \label{eq:diffusion}
    \tau_{diff} \sim \frac{H^2}{\kappa_{z,tracer}},
\end{equation}

\noindent where $H$ is given by the depth of the stratified layer as predicted by Equation~(\ref{eq:H}) (if it is smaller than the ocean depth) or the ocean depth if the stratified layer occupies the whole ocean. We choose $Q\approx 0.03$~W~m$^{-2}$ and $\left| \Delta T \right| \approx$2.0~K for a salinity of 8.5~g~kg$^{-1}$, following Section~\ref{subsec:stratification_theory}. Using the range for thermal and tracer vertical diffusivity estimated in Section~\ref{subsec:tidal}, we find the maximum mixing time scale to be $\tau_{diff}^{max} \approx 3.5 \times 10^5$~years when the turbulent diffusivity is $\kappa_{z,heat}=\kappa_{z,tracer}\approx1.4 \times 10^{-4}$~m$^2$~s$^{-1}$ such that the depth of the stratified layer is the same as the depth of the ocean, and the minimum time scale to be $\tau_{diff}^{min} \approx 250$~years when $\kappa_{z,heat}=\kappa_{z,tracer}=10^{-7}$~m$^2$~s$^{-1}$, i.e. the tracer diffusivity is enhanced by turbulent mixing to a similar value as the molecular thermal diffusivity. Vertical mixing through the stratified layer is therefore expected to take at least hundreds, and possibly up to hundreds of thousands of years. Notice that there may also be a buoyancy-driven circulation in the stratified layer \citep[due to inhomogeneities in the ice shell thickness and freshwater/salinity fluxes from melting/freezing of the ice shell; e.g.,][]{ashkenazy2020europa, kang2020differing, lobo2021pole, kang2021does}, which results in advective transport of the tracer. However, any such circulation would be diffusively controlled (i.e., a balance exists between vertical advection and diffusion), such that the advective time scale remains constrained by the diffusive time scale. The presence of a stratified layer in a low salinity ocean is therefore likely to greatly increase the vertical mixing time scale compared to a fully convective high salinity ocean.

\section{Numerical Simulations} \label{sec:simulations}
\subsection{Experimental Design} \label{subsec:experiments}

We perform numerical simulations using the Massachusetts Institute of Technology General Circulation Model (MITgcm) to solve the non-hydrostatic equations for a Boussinesq fluid in a rotating spherical shell, where all sphericity terms are preserved, including all components of the Coriolis force \citep[see][and APPENDIX \ref{si:model}]{adcroft2018mitgcm}. Radius, rotation rate and gravity are set to be the same as Enceladus (Table \ref{tb:parameters}), and vertical variation of gravity is also taken into consideration (Figure \ref{figS_gravity}(a), see APPENDIX \ref{si:model}).

\begin{figure}[b]
\centering
\includegraphics[width=0.7\linewidth]{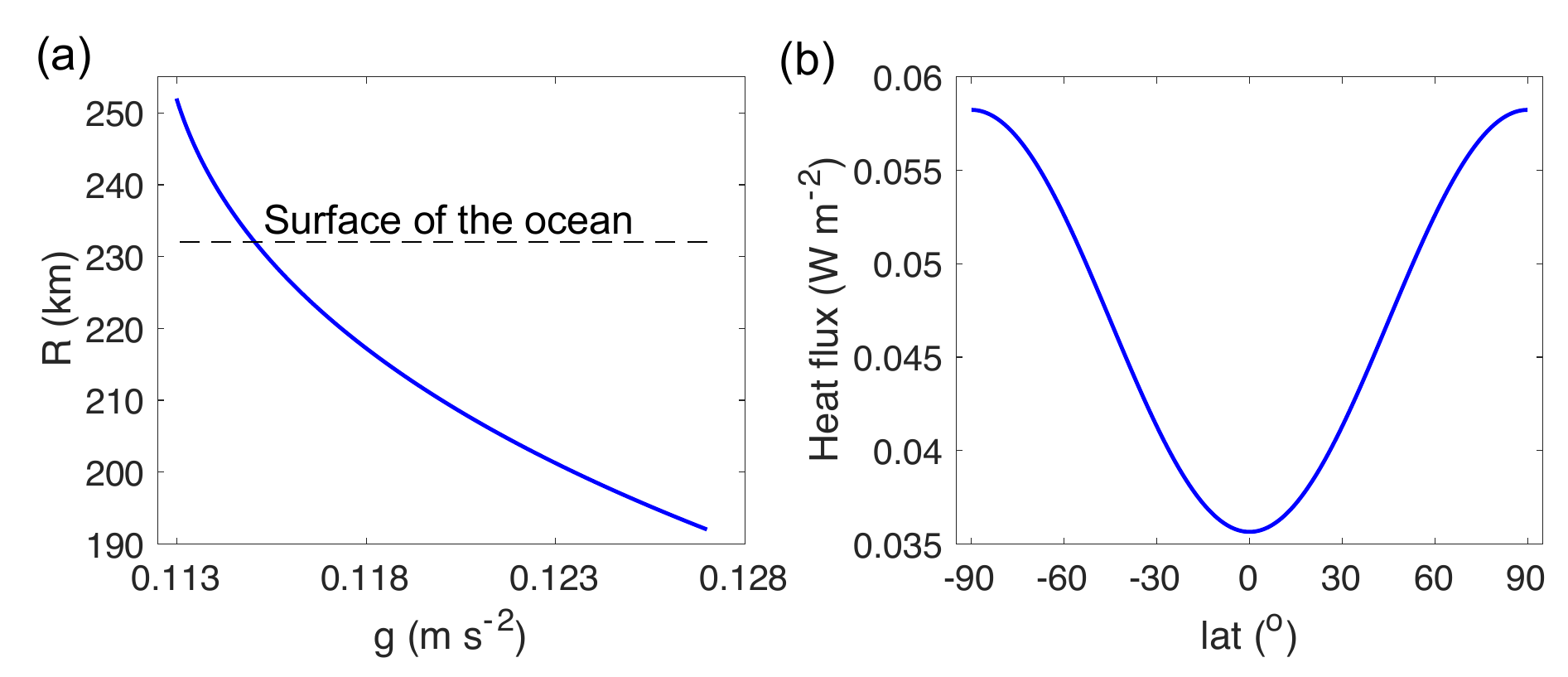}
\caption{Gravity and bottom heating pattern of simulations. (a) Vertical gravity profile. The black dashed line indicates the radius at the surface of the ocean. (b) Bottom heat flux. Note that there is no zonal variation in the heat flux.}
\label{figS_gravity}
\end{figure}

At the bottom of the ocean, we apply a fixed bottom heat flux pattern, following the zonal-mean tidal forcing pattern in \citet{choblet2017powering} with a total flux of 20~GW (Figure \ref{figS_gravity}(b), see APPENDIX \ref{si:model}). The bottom heat flux boundary condition is qualitatively similar to that in \citet{kang2020differing} and \citet{ashkenazy2020europa}, but differs from \citet{soderlund2019ocean} who applies fixed and uniform bottom and surface temperatures. The prescribed bottom heat flux is better constrained from tidal models and observations than the bottom temperature distribution and provides a key constraint on the energetics of the circulation \citep{jansen2016turbulent}. We apply a linear bottom drag with a drag coefficient $r_b\sim 10^{-4}$~m~s$^{-1}$ (such that the velocity in the bottom layer is relaxed to zero with a rate of 10$^{-7}$~s$^{-1}$). This value is likely to be unrealistically large, but we found a relatively large value to be needed to avoid slow spin-up of unrealistic mean flows. Our choice of a linear bottom drag is qualitatively similar to \citet{kang2020differing} (although they apply an even larger value of $r_b=2\times 10^{-3}$~m~s$^{-1}$), and is also practically similar to \citet{ashkenazy2020europa} who apply a no-slip bottom boundary condition together with vertical viscosity, but is different from \citet{soderlund2019ocean} who applies stress-free boundary conditions. We believe that some bottom drag is preferable as it provides a physical constraint on the bottom zonal flows.

\begin{deluxetable*}{cc}
\tablenum{1}
\tablecaption{Parameters for Enceladus simulations \label{tb:parameters}}
\tablewidth{0pt}
\tablehead{
\colhead{Parameters} & \colhead{Value}
}
\startdata
Surface gravity $g(r_s)$ & 0.113 m s$^{-2}$ \\
Ice shell surface radius $r_s$ & 252 km \\
Ocean surface radius $r_o$ & 232 km \\
Ocean bottom radius $r_c$ & 192 km \\
Rotation rate $\Omega$ & $5.31\times 10^{-5}$ s$^{-1}$ \\
Prandtl number $Pr=\nu/\kappa$ & 10 \\
\enddata
\tablecomments{The values are taken from \citet{vcadek2019long} with some simplifications.}
\end{deluxetable*}

At the top model layer, we apply a linear restoring of temperature towards the freezing point $T_f$ ($-2~^\circ$C for salinity of 35~g~kg$^{-1}$ and $-0.6~^\circ$C for salinity of 8.5~g~kg$^{-1}$) with a restoring time of 1~month (30~days). A 30 day restoring time scale is fast compared to the typical advective time scale, so that this boundary condition effectively amounts to a fixed temperature boundary condition. Our upper boundary condition can implicitly capture the heat exchange between the surface global ice shell and the ocean, but the fresh water flux and brine rejection associated with melting and freezing are not included in our simulations. Salinity changes associated with freezing and melting at the surface of the ocean can change the density of sea water, thus further affecting ocean circulation. A pole-to-equator overturning circulation driven by salinity gradients, which in turn are caused by ice formation at low-latitudes and melting at the pole, has been suggested in \citet{lobo2021pole} and has also been found in coupled ice-ocean simulations using the MITgcm \citep{ashkenazy2020europa, kang2020differing, kang2021does}. Although freezing and melting at the bottom of the ice shell is likely to affect the ocean circulation, it is important to note that such forcing cannot energetically drive a circulation unless melting (which reduces the density) happens at a higher pressure (i.e. greater depth) than freezing \citep[e.g.][]{wunsch2004vertical}. Instead, assuming that the ice shell is in equilibrium, any melting or freezing of the ice\footnote{Melting and freezing of the ice is expected to be affected by tidal dissipation in the ice shell \citep[e.g.][]{kite2016sustained, souvcek2019tidal, beuthe2019enceladus, kang2020spontaneous}, the ``ice pump'' effect associated with the freezing point depression at increasing pressure \citep{lewis1986ice}, as well as ocean dynamics, which modulate heat flux to the ice shell. The mass budget of the ice shell, however, has to close independently of these mechanisms and, if the ice shell is in equilibrium, directly relates the net freezing/melting rate to the divergence/convergence of the ice flow.} must be balanced by the viscous ice flow \citep[e.g.,][]{lefevre2014structure, corlies2017titan, ashkenazy2018dynamics, vcadek2019long}, which tends to advect ice from thicker regions to thinner regions. As a result, melting needs to occur where the ice shell is thin (i.e. at small depth), while freezing would have to occur where the ice shell is thick. The salinity forcing then cannot drive a circulation but instead acts to stabilize the stratification in regions where the ice shell is relatively thin. The energy source for the global overturning circulation in \citet{lobo2021pole}, and the ``salt-driven'' circulation in the simulations of \citet{ashkenazy2020europa}, \citet{kang2020differing}, and \citet{kang2021does}, comes from parameterized turbulent vertical mixing where the ocean is stably stratified. Whether a strong ``salt-driven'' circulation is possible on Enceladus thus remains an open question that directly ties in with the question of how much turbulent kinetic energy is available for vertical mixing, which is here left for future work.

We simulate a 40~km deep ocean over a zonal range of 15$^\circ$ with zonally periodic boundary conditions and a meridional range from 85.5$^\circ$S to 85.5$^\circ$N, with free-slip, no-normal flow conditions at the meridional boundaries. The longitudinal extent of the domain is limited to save computational resources and the poles are masked with land to ensure numerical stability. The vertical resolution is 1000~m. The horizontal resolution is 1$^\circ$ in the zonal and 0.95$^\circ$ in the meridional direction in most simulations, which is around 4~km $\times$ 4~km near the equatorial surface. This horizontal resolution is not sufficient to adequately resolve single convective columns. The expected minimum horizontal scale of the convective columns can be estimated using the length scale where rotation becomes important: $l_r \sim B^{1/2}f^{-3/2} \approx 0.2$~m \citep{jones1993convection}, which is many times smaller than the horizontal resolution. This indicates that in our simulations, convection is affected by the resolution and parameterized turbulent diffusivities and viscosities, a situation that is unavoidable in global-scale simulations, where resolutions of O(0.1~m) are computationally impossible to achieve. In order to provide at least some insight into how resolution can affect the simulation results, we perform one additional simulation with a finer horizontal resolution of 0.5$^\circ$ in the zonal and 0.475$^\circ$ in the meridional direction.

\begin{deluxetable*}{ccccc}
\tablenum{2}
\tablecaption{Simulation Settings\label{tb:experiments}}
\tablewidth{0pt}
\tablehead{
\colhead{Case} & \colhead{Salinity} &  \colhead{$\kappa_{h,heat}$} & \colhead{$\kappa_{z,heat}$} & \colhead{Horizontal} \\
\colhead{Descriptions} & \colhead{(g kg$^{-1}$)} &  \colhead{(m$^2$ s$^{-1}$)} & \colhead{(m$^2$ s$^{-1}$)} & \colhead{Resolution}
}
\startdata
\textit{HSaniso} & 35 & 0.25 & 5$\times$10$^{-5}$ & 1$^\circ \times$0.95$^\circ$ \\
\textit{HSiso} & 35 & 0.25 & 0.25 & 1$^\circ \times$0.95$^\circ$ \\
\textit{HSiso05} & 35 & 0.1 & 0.1 & 0.5$^\circ \times$0.475$^\circ$ \\
\textit{LSaniso} & 8.5 & 0.25 & 5$\times$10$^{-5}$ & 1$^\circ \times$0.95$^\circ$ \\
\enddata
\end{deluxetable*}

The horizontal turbulent diffusivity applied in our models is 0.25~m$^2$~s$^{-1}$ in our $1^\circ \times 0.95^\circ$ simulations, chosen for numerical stability. In the high horizontal resolution simulation, this value is decreased to 0.1~m$^2$~s$^{-1}$ following Kolmogorov scaling (see APPENDIX~\ref{si:model}). The turbulent vertical diffusivity is likely to be small in the stratified layer of the low salinity ocean, as discussed in Section~\ref{subsec:tidal}. For the low salinity ocean we therefore use an anisotropic diffusion with a smaller vertical diffusivity set to $\kappa_z= 5 \times 10^{-5}~\mathrm{m^2~s^{-1}}$. This value is similar to the vertical diffusivity in Earth's ocean and was chosen here to ensure numerical stability and to be able to explicitly resolve the stratified layer. The turbulent vertical diffusivity in the high salinity ocean is not constrained by the energetic argument discussed in Section~\ref{subsec:tidal}, as no energy is required to mix the unstratified ocean. The horizontal and vertical scale of the grid in our simulations are of similar order (10$^3$~m), so that an isotropic diffusion regime is plausible. However, due to the effect of gravity and rotation, highly anisotropic turbulent diffusion may also be justified. We have therefore performed simulations with both isotropic and anisotropic diffusion for the high salinity ocean. All viscosities are set by fixing $Pr=10$.

We carry out 4 simulations to examine the influence of different factors on ocean circulation (Table \ref{tb:experiments}). In order to examine the role of salinity, we carry out two simulations, \textit{HSaniso} and \textit{LSaniso}, with two different salinities: $35~\mathrm{g~kg^{-1}}$ for the high salinity case (similar to Earth's ocean) and $8.5~\mathrm{g~kg^{-1}}$ for the low salinity case \citep{glein2018geochemistry}. We perform simulations with both isotropic and anisotropic diffusion (\textit{HSiso} versus \textit{HSaniso}) in high salinity oceans to test the robustness of our results. The isotropic diffusion is consistent with \citet{soderlund2019ocean}, and the anisotropic diffusion is consistent with \citet{kang2020differing}, as well as with our low salinity simulation. In order to examine the influence of resolution, we set up a simulation with doubled horizontal resolution (\textit{HSiso05} compared with \textit{HSiso}). We only test the effect of resolution in the high salinity ocean because the low salinity ocean takes a much longer time to reach an equilibrium state, due to the diffusive adjustment of the stratified layer, which makes the simulation computationally very expensive. All simulations are integrated to a near equilibrium state in which the energy imbalance is less than 2\%. The presented results are 10-year-averages for high salinity cases and 250-year-averages for the low salinity case after this equilibrium has been reached. The longer averaging time in the low salinity case was chosen to account for a much larger low-frequency variability in this simulation.

To study tracer mixing processes in the ocean, we carry out three tracer simulations initialized from the equilibrium states of \textit{HSaniso}, \textit{HSiso} and \textit{LSaniso}. We initialize two passive tracers (i.e. with no effect on ocean density and thus dynamics) at the bottom of the ocean at 0$^\circ$ and 60$^\circ$S to study the evolution of tracer concentration. The turbulent diffusivity for the tracers is set to be the same as the thermal turbulent diffusivity. Each tracer simulation is run for 1500~years, and 10-year-averages are used for analysis.

\subsection{Simulation Results} \label{subsec:simulation_result}
\subsubsection{Vertical Stratification and Circulation} \label{subsubsec:sim_strat}

\begin{figure}[b]
\centering
\includegraphics[width=0.85\linewidth]{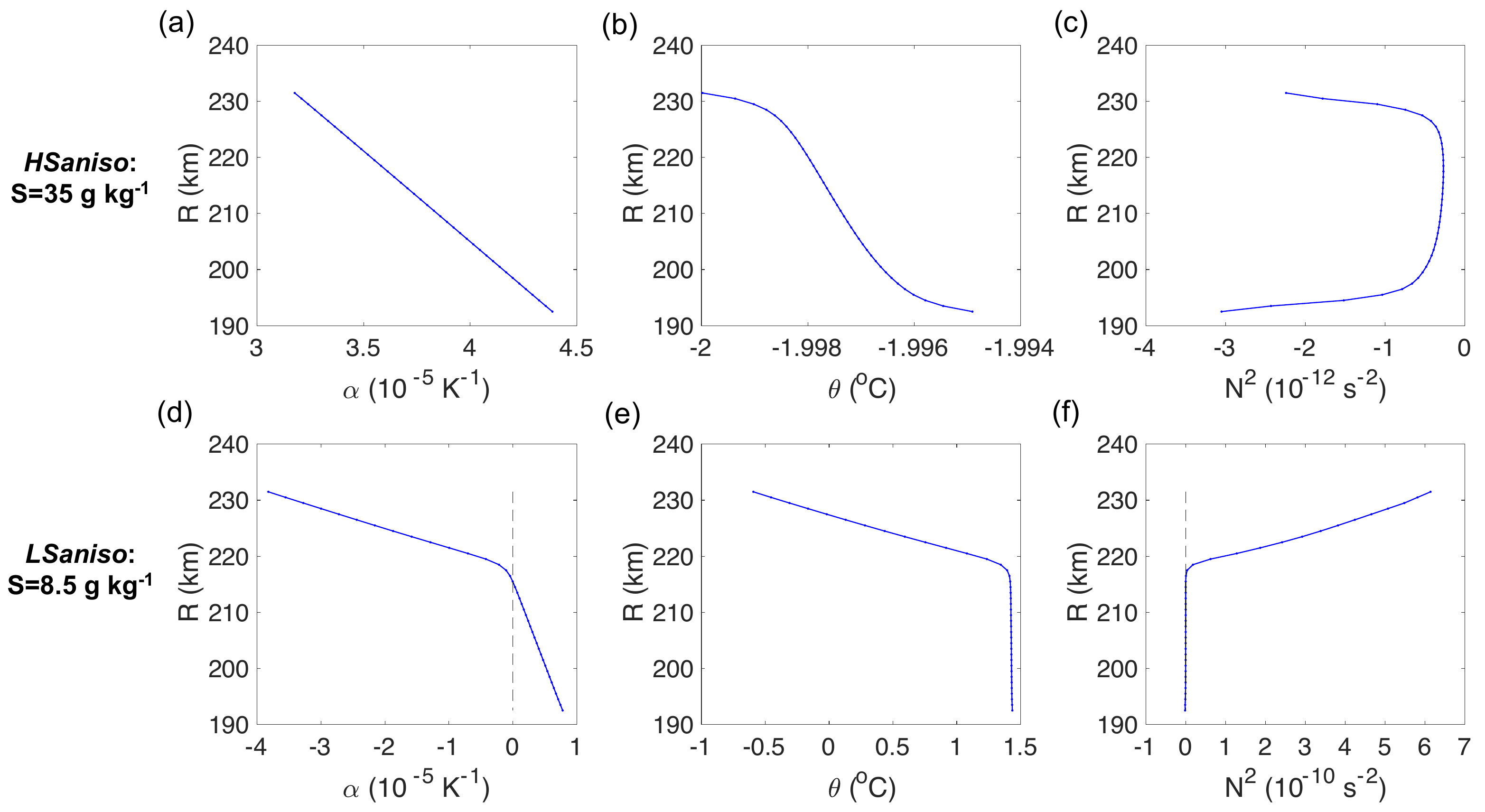}
\caption{Vertical stratification of the high and low salinity ocean (top row and bottom row, respectively). From left to right are the horizontally averaged vertical profile of thermal expansivity $\alpha$, potential temperature $\theta$ and stratification $N^2 \approx \alpha g \partial \theta / \partial z$. The black dashed line marks the zero-line.}
\label{fig_stratification}
\end{figure}

\begin{figure}[ht]
\centering
\includegraphics[width=0.84\linewidth]{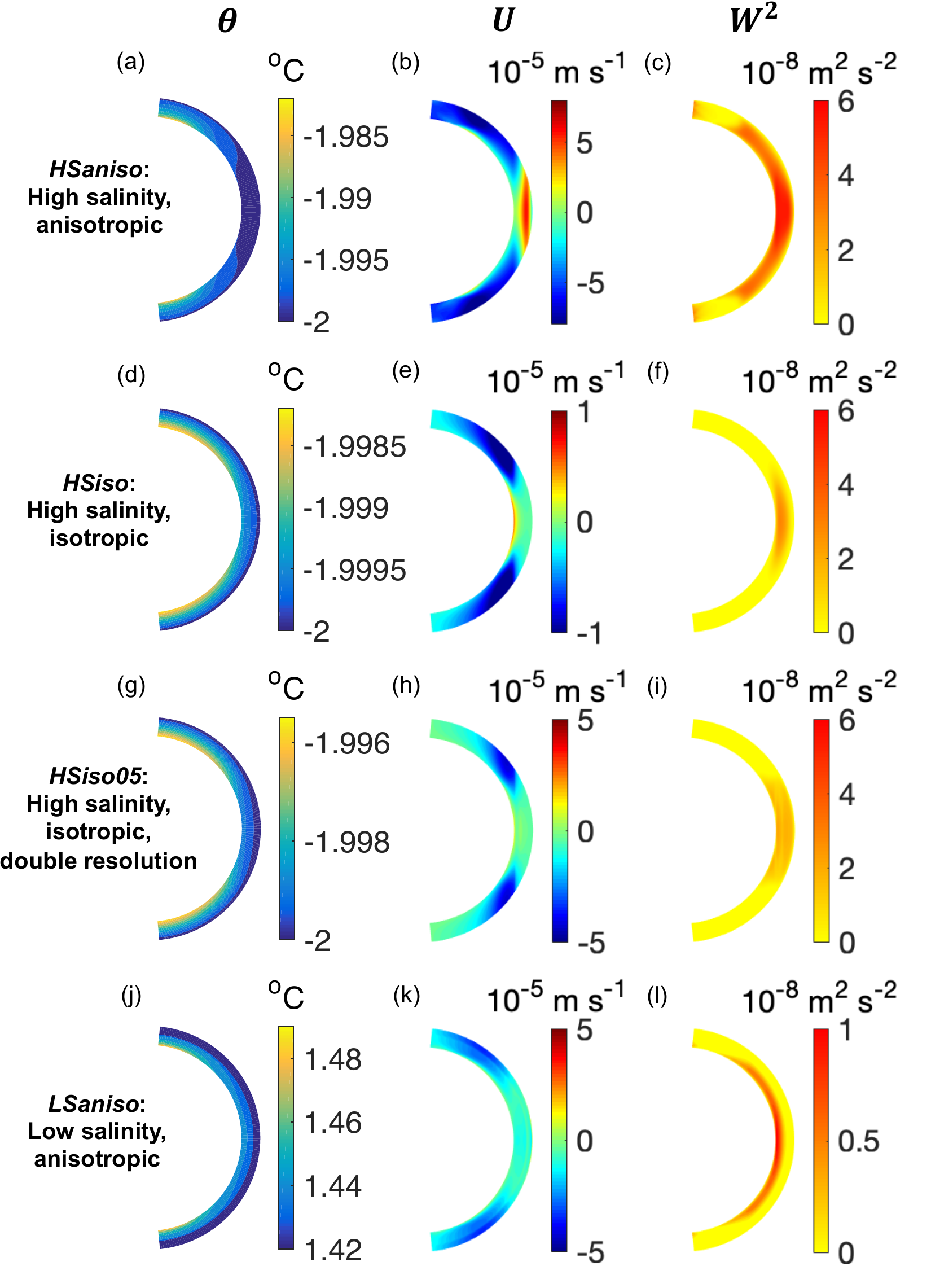}
\caption{Ocean potential temperature and flow fields in the equilibrium state of different simulations. From left to right, the three columns show time-averaged zonal-mean fields of potential temperature $\theta$, zonal velocity $U$ and vertical kinetic energy $W^2$, respectively. Note that the colorbar is saturated in panel (j). The surface temperature in \textit{LSaniso} is very low (-0.6~$^\circ$C) and increases with depth linearly in the stratified layer, as shown in Figure \ref{fig_stratification}(e).}
\label{fig_basic}
\end{figure}

In our high salinity simulations, the ocean stratification is weakly convectively unstable at all depths (Figure~\ref{fig_stratification}(a)-(c), and Figure~\ref{fig_basic}(a), (d) \& (g)). In \textit{HSiso} and \textit{HSiso05}, the resolved convection at mid- and high-latitudes is, however, weak (Figure~\ref{fig_basic}(f) \& (i)). This result can be understood by noting that the large vertical diffusivity and viscosity in these two cases leads to a relatively low Rayleigh number ($Ra=\alpha g \delta T D^3 / \nu \kappa$ where $\delta T$ is the temperature contrast between the bottom and surface of the ocean and $\nu$ is viscosity) of $Ra \approx 1 \times 10^6$ in \textit{HSiso} and $Ra \approx 1.5 \times 10^7$ in \textit{HSiso05}. Importantly, the Rayleigh number in these two cases is smaller than the critical value that has been suggested for rotating convection: $Ra_S=8.7E^{-4/3}$ where $E=\nu /(2\Omega D^2)$ is the Ekman number \citep{cheng2018heuristic}, which is $E\approx1.5 \times 10^{-5}$ in \textit{HSiso} and $E\approx6 \times 10^{-6}$ in \textit{HSiso05}. The critical Rayleigh number is thus $Ra_S \approx 2 \times 10^7$ in \textit{HSiso} and $Ra_S \approx 8 \times 10^7$ in \textit{HSiso05}, which is larger than the respective Rayleigh numbers. As a result, radial heat flux is dominated by the parameterized turbulent diffusion.

Temperature gradients are very small in all high salinity simulations, with the whole ocean near the freezing point. The bottom to surface buoyancy contrast is $\alpha g \delta T \approx 10^{-7}$~m~s$^{-2}$ in \textit{HSaniso} and $\alpha g \delta T \approx 10^{-8}$~m~s$^{-2}$ in \textit{HSiso} and \textit{HSiso05} (Figure~\ref{fig_basic}(a), (d) \& (g)). For comparison, the prescribed buoyancy contrast in \citet{soderlund2019ocean} is around $10^{-5}$~m~s$^{-2}$ for Enceladus, which is at least two orders of magnitude larger than our result, consistent with a radial heat flux in the simulations of \citet{soderlund2019ocean} that is many orders of magnitude larger than the observed surface heat loss. 

The zonal flow at mid-latitudes is constrained by the temperature structure via the thermal wind relationship, in particular the equator-to-pole temperature gradient, with warmer temperatures at the poles and colder temperatures at low latitudes, which leads to a negative (westward) vertical current shear at mid-latitudes (second colomn in Figure~\ref{fig_basic}). In \textit{HSaniso}, there exists superrotation in the equatorial upper ocean (Figure~\ref{fig_basic}(b)), while such superrotation does not exist in \textit{HSiso} and \textit{HSiso05} (Figure~\ref{fig_basic}(e) \& (h)), suggesting that the equatorial dynamics are sensitive to poorly constrained simulation parameters. The mechanism for the superrotation is associated with upward eddy momentum fluxes, consistent with previous studies \citep[see APPENDIX~\ref{si:superrotation} for a more detailed discussion of the mechanism for superrotation]{aurnou2001strong, kaspi2008turbulent, ashkenazy2020europa}. Note that the simulation results of \textit{HSiso} and \textit{HSiso05} with different resolutions are qualitatively similar but differ quantitatively (compare the 2$^{nd}$ and 3$^{rd}$ rows in Figure~\ref{fig_basic}).

As predicted, there are two different vertical layers in the low salinity ocean (Figure~\ref{fig_stratification}(d)-(f)). The bottom layer is a convective layer, which is slightly negatively stratified with positive thermal expansivity and small temperature variation around the critical point $T_c$ (Figure~\ref{fig_stratification}(e) \& Figure~\ref{fig_basic}(j)). In the time-averaged equilibrium state, convection in the convective layer in the low salinity ocean (Figure~\ref{fig_basic}(l)) is weaker than in the high salinity ocean (Figure~\ref{fig_basic}(c)), as expected due to the smaller thermal expansivity and thus smaller buoyancy input. The upper layer is the stratified layer with stable stratification and negative thermal expansivity, so that convection is blocked at the bottom of this layer (Figure~\ref{fig_basic}(l)). The vertical temperature gradient is relatively large in this layer, with a profile that is linearly increasing from the freezing point $T_f$ at the surface of the ocean to the critical temperature $T_c$ at the interface between the two layers. We can apply Equation (\ref{eq:H}) to estimate the expected depth of the stratified layer. In our simulation, the temperature variation is $\left| \Delta T \right| \approx$2.0~K, the vertical heat flux is $Q\approx$0.03~W~m$^{-2}$, and the vertical thermal diffusivity is $\kappa_{z,heat} = 5\times$10$^{-5}$~m$^2$~s$^{-1}$, such that the predicted depth of the stratified layer is around 14~km. This estimated depth matches well with the simulation results (Figure~\ref{fig_stratification}(d)-(f)). Notice that the specific thickness of the stratified layer is sensitive to our assumption for the vertical diffusivity, which is poorly constrained. However, the fact that the numerical simulations support the theoretical predictions of Section~\ref{sec:theory} lends support to the more general scaling arguments discussed there.

\subsubsection{Ocean Heat Transport} \label{subsubsec:heat_flux}

\begin{figure}[t]
\centering
\includegraphics[width=0.634\linewidth]{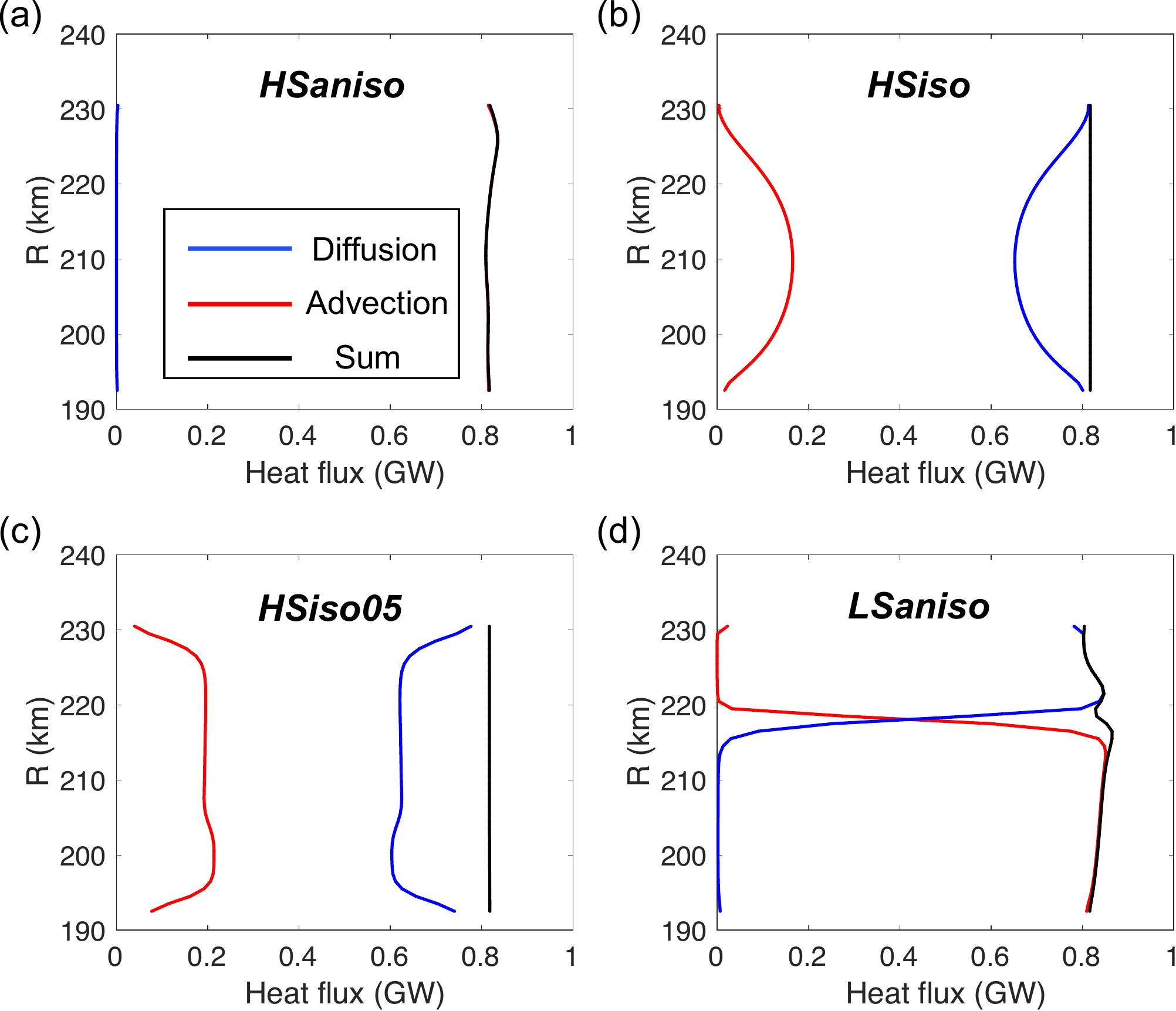}
\caption{Vertical heat flux in the simulations. Panels show horizontally integrated vertical heat flux as a function of depth in \textit{HSaniso} (a), \textit{HSiso} (b), \textit{HSiso05} (c), and \textit{LSaniso} (d). Blue lines indicate the heat flux due to parameterized turbulent diffusion, red lines indicate the heat flux due to resolved advection, and black lines indicate the total heat flux.}
\label{fig_heatflux}
\end{figure}

The mechanisms governing ocean heat transport vary widely across our simulations. With anisotropic diffusivity and viscosity (\textit{HSaniso}) the vertical heat flux in the high salinity simulation is dominated by explicitly resolved convection (Figure~\ref{fig_heatflux}(a)). With isotropic diffusivity and viscosity (\textit{HSiso} and \textit{HSiso05}) instead, the vertical heat flux is dominated by parameterized turbulent diffusion (Figure~\ref{fig_heatflux}(b)~\&~(c)) due to the large vertical diffusivity.

\begin{figure}[b]
\centering
\includegraphics[width=0.636\linewidth]{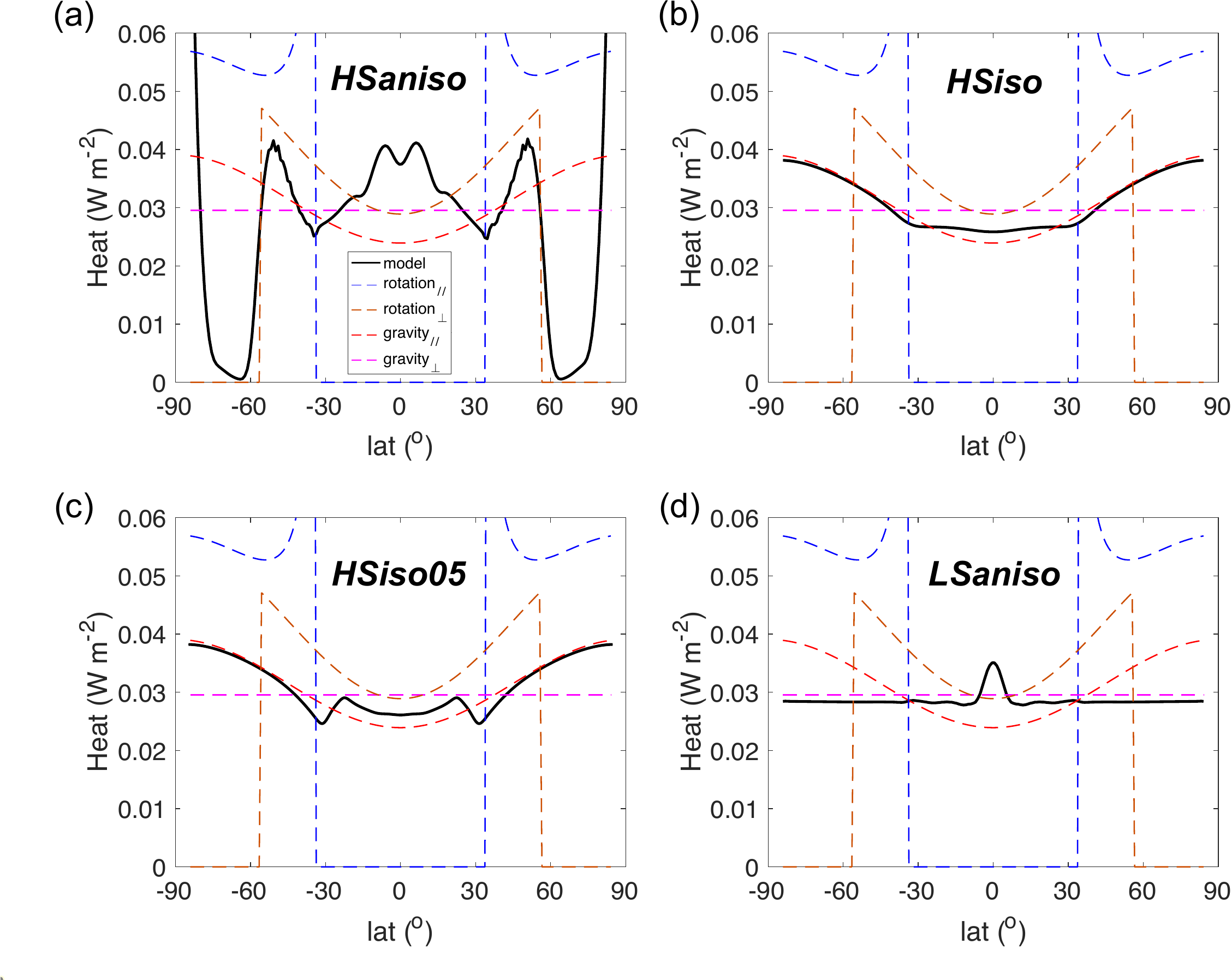}
\caption{Ocean surface heat flux in simulations compared with four conceptual models. Black lines indicate the surface heat flux in the simulations. Dashed lines indicate the surface heat flux in the four conceptual models, where blue lines indicate the ``Parallel to Rotation'' model, brown lines indicate the ``Perpendicular to Rotation'' model, red lines indicate the ``Parallel to Gravity'' model, and pink lines indicate the ``Perpendicular to Gravity'' model.}
\label{fig_surfHeat}
\end{figure}

The choice of anisotropic versus isotropic diffusivity and viscosity also affects the heat flux at the ice-ocean interface, and by comparing the ocean surface heat flux from simulation results with the conceptual models in Section~\ref{subsec:HeatModel} we find that different aspects appear to play a role in different configurations. In \textit{HSaniso}, the peaks at around 60$^\circ$ match well with the ``Perpendicular to Rotation'' model, indicating the important role of heat transport by the horizontal currents around Taylor columns. Heat fluxes are further enhanced in the equatorial region, associated with particularly strong ``equatorial rolls'', which have previously been discussed in \citet{kang2020differing}. Notice that the strong peaks near the poles are likely to be associated with the artificial boundaries at $85^\circ$N/S (Figure~\ref{fig_surfHeat}(a)). Although the result that vertical heat transport is dominated by convection and associated Taylor columns is likely to be robust, the specific patterns of the convection are likely to depend on model resolution and poorly constrained parameters, and the resolution remains inadequate to resolve the natural scale of rotating convection (c.f. Section~\ref{subsec:experiments}). In particular, the highly anisotropic viscosity and diffusion coefficients clearly affect the dynamics significantly. Although unresolved turbulence is likely to be anisotropic, even in a high salinity scenario, both the strength and alignment of the anisotropy remain unknown.

\begin{figure}[b]
\centering
\includegraphics[width=1.0\linewidth]{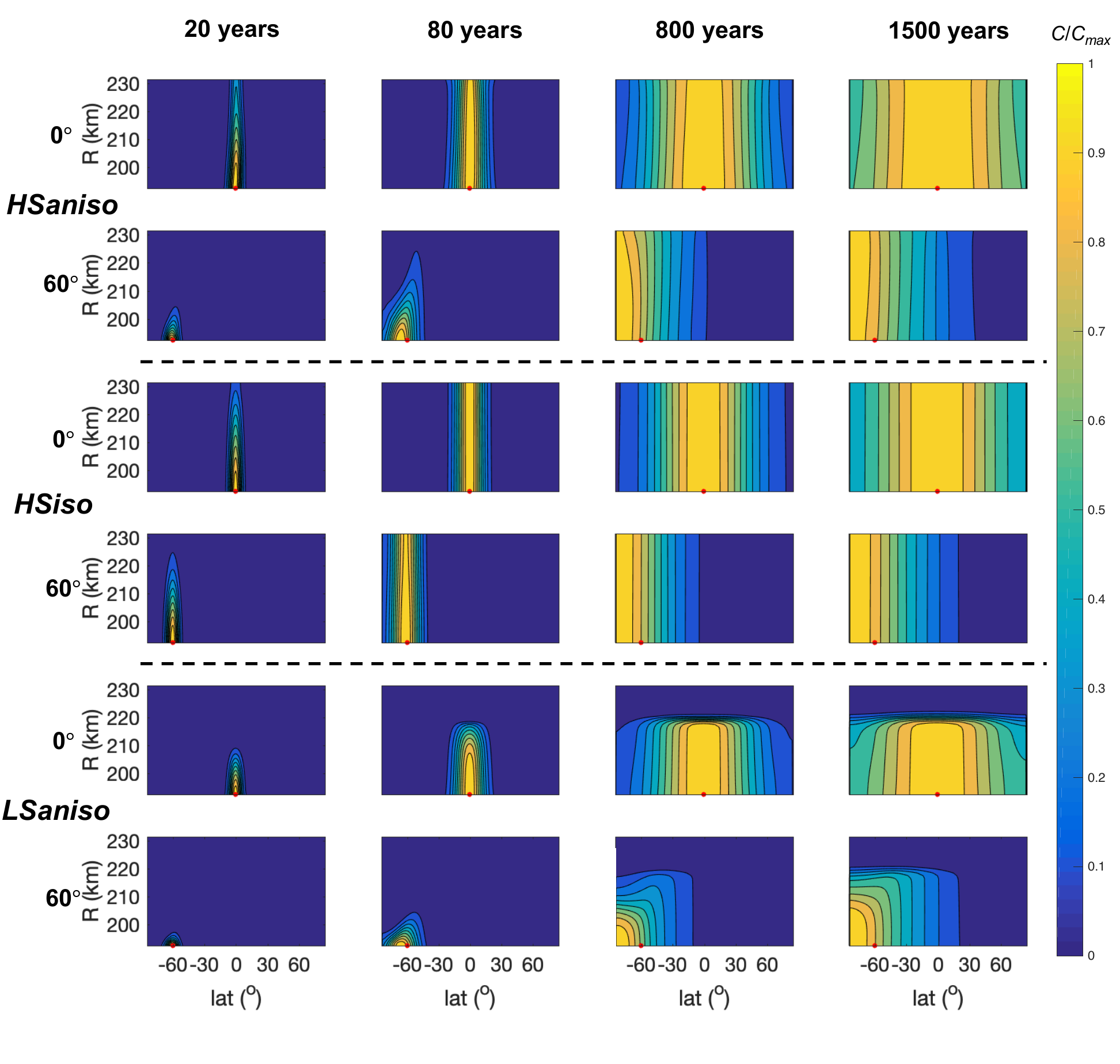}
\caption{Zonally-integrated tracer concentration as a function of latitude and depth. The concentration is normalized by the maximum value in each plot. From top to bottom, rows show simulation results in which the tracer is initialized at the ocean floor at 0$^\circ$ and 60$^\circ$S (the red dots in each panel) in \textit{HSaniso}, \textit{HSiso}, and \textit{LSaniso}, respectively. From left to right, columns show 10-year-averages after 20, 80, 800 and 1500~years, respectively.}
\label{fig_tracer}
\end{figure}

In \textit{HSiso} and \textit{HSiso05}, the surface heat fluxes are very close to the  ``Parallel to Gravity'' model (Figure~\ref{fig_surfHeat}(b) \& (c)), consistent with the result that vertical heat flux is dominated by parameterized isotropic turbulent diffusion. The heat flux at the equator is somewhat larger than predicted by the ``Parallel to Gravity'' model because it is enhanced by heat transport through the ``equatorial rolls''. In \textit{HSiso05}, there are small peaks outside the tangent cylinder, which are caused by convection (c.f. Figure~\ref{fig_basic}(i)). These peaks qualitatively resemble those predicted by the ``Parallel to Rotation'' model, potentially indicating the effect of slantwise convection, although their amplitude is very weak, due to the dominant role of diffusion. The surface heat flux patterns in these two simulations are qualitatively similar to the heat flux estimated in \citet{vcadek2019long} based on Cassini observations, with the same order of magnitude (O(10~mW~m$^{-2}$)) and heat loss rates increasing with latitude, although the spatial variations of the heat fluxes are smaller in our simulations.

In the low salinity case, \textit{LSaniso}, the vertical heat transport is dominated by convection in the bottom convective layer, but is dominated by diffusion in the upper stratified layer (Figure~\ref{fig_heatflux}(d)). The surface heat flux is almost uniform, and well approximated by the ``Perpendicular to Gravity'' model (Figure~\ref{fig_surfHeat}(d)). The approximately uniform surface heat flux is a result of the relatively strong horizontal mixing in the stably stratified layer where the horizontal diffusivity is much larger than the vertical diffusivity. The peak at the equator is associated with grid-scale vertical advection, which is likely to be an artifact of numerical instability.

\subsubsection{Ocean Tracer Mixing} \label{subsubsec:tracer}

The three tracer simulations illustrate the mixing processes in the ocean and the time it takes for constituents to be mixed from the bottom to the surface of the ocean. In the high salinity ocean, tracers can reach the surface of the ocean within tens of years. The convective mixing time scale is within an order of magnitude but somewhat shorter than the estimate according to Equation~(\ref{eq:convection}) because the scaling underestimates the vertical velocities in the simulations. Horizontal mixing across the hemisphere takes around 1000 years, while global mixing across the equator remains incomplete by the end of our 1500-year simulation (Figure~\ref{fig_tracer}). The mixing time scales are only moderately dependent on the choice of isotropic versus anisotropic diffusion. Note that the poleward drift of the highest concentration center and the tilting pattern arise due to spherical geometry effects on the diffusive transport as well as advection by meridional currents.

In the low salinity simulation, \textit{LSaniso}, the mixing time scale in the bottom convective layer is similar to that in the high salinity simulations. However, the tracers do not penetrate significantly into the stratified layer, even after 1500~years (Figure \ref{fig_tracer}). The vertical mixing time scale in the stratified layer can be estimated using Equation~(\ref{eq:diffusion}). In our simulation, $\kappa_{z,heat}=\kappa_{z,tracer}=5\times$10$^{-5}$~m$^2$~s$^{-1}$, so that $\tau_{diff} \approx 4 \times 10^{12}$ s $\approx 1 \times 10^5$~years. This is consistent with the simulation results showing little tracer penetration into the stratified layer after 1500~years. Note that this specific time scale is sensitive to our assumption for the vertical diffusivity, which is poorly constrained. However, the fact that the numerical simulations are consistent with the scaling arguments in Section~\ref{sec:theory}, increases our confidence in the theoretical prediction that the vertical mixing time scale in the stratified layer would be between hundreds and hundreds of thousands of years, depending on the assumed turbulent vertical diffusivity.

\section{Conclusion} \label{sec:conclusion}

We find that the salinity of Enceladus' ocean fundamentally determines the vertical stratification and circulation of the ocean, thereby affecting the heat and tracer transport from the rocky core to the surface ice shell. If salinity is high, the ocean is unstratified and is dominated by convection over the whole depth. Tracers can be transported from the bottom to the surface within tens of years. In a low salinity ocean (below a critical point of about 20~g~kg$^{-1}$), a stably stratified layer exists in the upper ocean, whose depth may be anywhere from tens of meters to the full depth of the ocean, depending on assumptions about the strength of tidally (and/or librationally) driven turbulence that can contribute to mixing of the stratified layer. The stratified layer strongly suppresses vertical mixing, leading to a vertical mixing time scale between hundreds and hundreds of thousands of years. In the presence of significant horizontal mixing, the stratified layer further leads to horizontally homogeneous heat transport to the ice shell, independent of any spatial structure in the heat flux underneath. More detailed simulation results are sensitive to model resolution, as well as the magnitudes of the assumed turbulent viscosities and diffusivities, which remain poorly constrained.

\begin{figure}[t]
\centering
\includegraphics[width=0.4\linewidth]{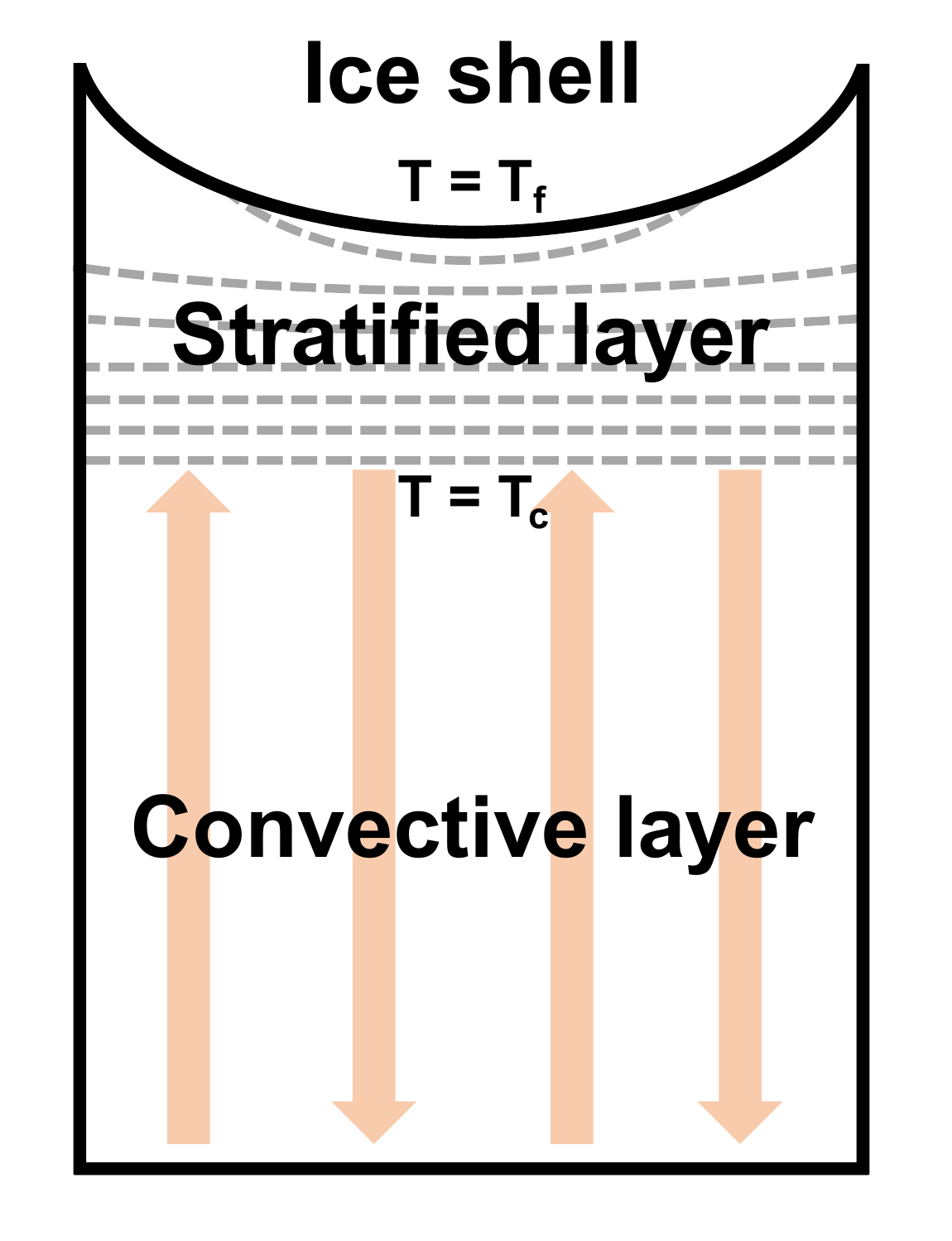}
\caption{Sketch of the expected temperature structure of the low salinity ocean with varying ice shell thickness. Grey dashed lines indicate the contour lines of the temperature, which increases with depth in the stratified layer. The temperature is fixed near the local freezing point $T_f$ at the ice-ocean interface and critical temperature $T_c$ at the interface between the stratified layer and the convective layer. The temperature at the surface of the ocean is colder at the equator but warmer at the poles because the freezing point $T_f$ decreases with pressure, but this temperature contrast is small compared to the temperature contrast across the stratified layer.}
\label{fig_sketch}
\end{figure}

Our simulations are based on an ocean-only model, hence the effect of freezing and melting on salinity as well as variations in the ice thickness are not included. When the ice thickness varies, the temperature at the ice-ocean interface (equal to the local freezing point $T_f$) decreases with depth. However, this temperature variation is around 0.1~K, assuming an ice shell thickness variation of around 20~km \citep[c.f. Figure 2 in][]{kang2021does}, which is much smaller than the difference between the freezing point $T_f$ and the critical temperature $T_c$ in a low salinity ocean ($T_c-T_f>$1~K at a salinity of 8.5~g~kg$^{-1}$). Spatial variations in the temperature contrast across the stratified layer and associated variations in layer thickness are hence expected to be relatively small. We moreover expect that horizontal variations in the layer thickness and height would be kept small by baroclinic instability and circulations that would arise in the presence of significant horizontal density gradients. A crude sketch of the temperature structure that may be expected in the presence of varying ice thickness is shown in Figure~\ref{fig_sketch}, although we note that this prediction is speculative and needs to be tested. Future work is required to investigate dynamics of the ocean on Enceladus when the ocean is coupled with an ice shell of spatially varying thickness. Care must be taken in any such study to investigate the sensitivity of the results to model parameters and resolution, which are always based on compromises enforced by limited computational resources.

Our results indicate a contradiction in estimates of salinity and vertical mixing time scale of Enceladus' ocean in previous studies. Studies on the geochemistry of Enceladus' ocean have generally indicated a relatively low salinity, likely less than 20~g~kg$^{-1}$ \citep[e.g.][]{postberg2009sodium, hsu2015ongoing, glein2018geochemistry}. In this case we expect a stably stratified layer to form and result in a vertical mixing time scale between the rocky sea floor and the ice shell of at least hundreds of years. However, the detection of silica nanoparticles in plumes has been argued to set an upper limit of several years on the mixing time scale, based on the size and growth rate of the particles \citep{hsu2015ongoing}. Our results suggest that these inferences are not compatible with each other, indicating that Enceladus' ocean is either saltier than previously suggested or the interpretation of silica nanoparticles needs to be reconsidered. One possible explanation is that melting of ice in the polar regions due to stronger tidal heating \citep{kang2020differing} may freshen the water expelled in the plumes, thus possibly leading to an underestimate of the bulk ocean salinity by observations of plume constituents. However, more work is needed to establish a consistent picture of Enceladus' ocean salinity and mixing time scale.

\acknowledgments
We are grateful to Jun Yang, Edwin S. Kite, Wanying Kang, Dorian S. Abbot, Mikael Beuthe and two anonymous reviewers for helpful discussions and comments. Y.Z. thanks the Department of Atmospheric and Oceanic Sciences and School of Physics at Peking University for the financial support during the summer exchange program at University of Chicago. This work was completed with resources provided by the University of Chicago Research Computing Center.

\appendix

\section{Calculation of Surface Heat Fluxes in Four Conceptual Models} \label{si:heat}

In this part, we will explain the calculation of surface heat fluxes in the four conceptual models shown in Figure~\ref{fig_HeatModel}.

\begin{figure}[b]
\centering
\includegraphics[width=0.9\linewidth]{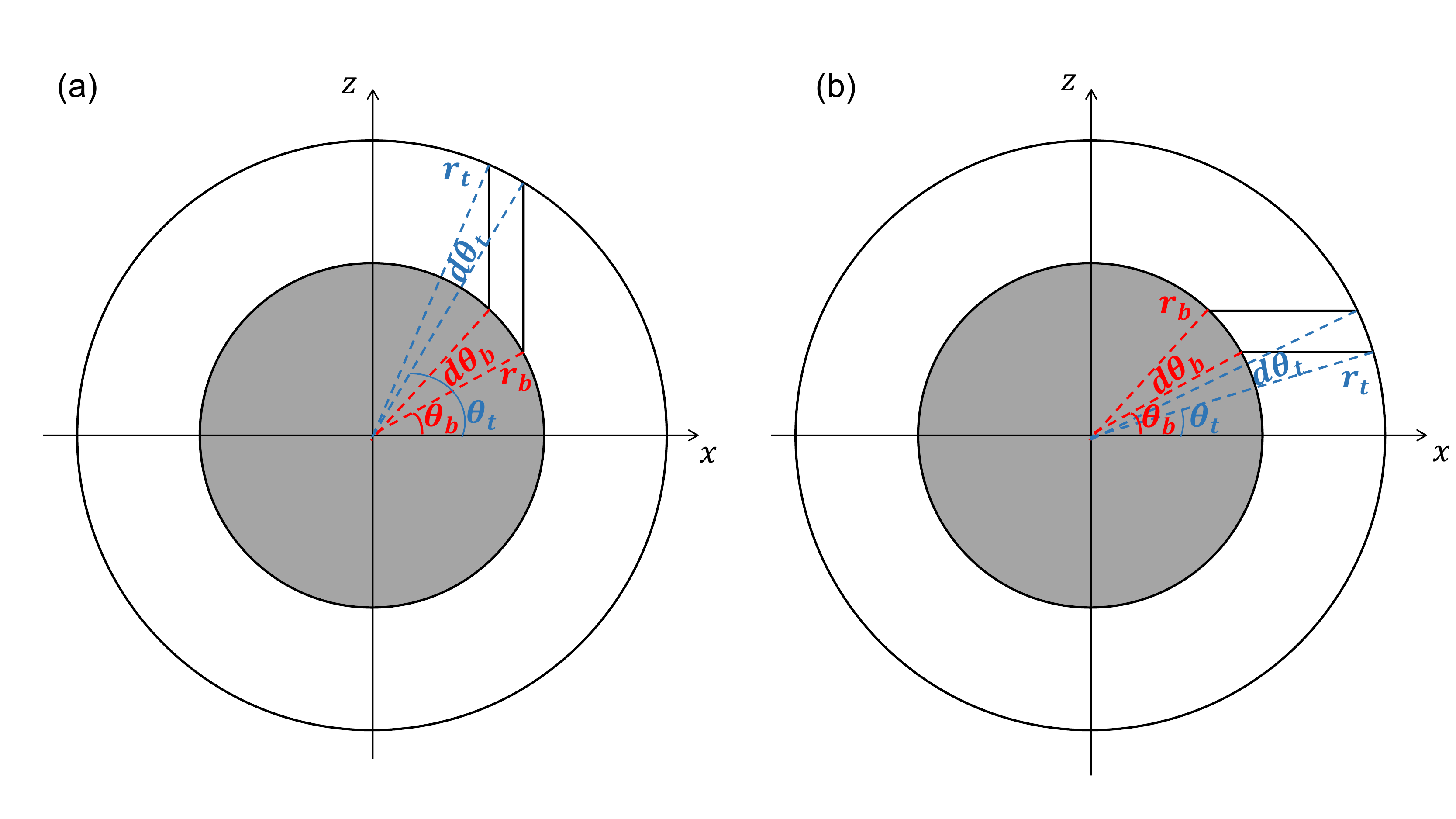}
\caption{Schematic for the ``Parallel to Rotation'' model (a) and the ``Perpendicular to Rotation'' model (b). The rotation axis is the $z$-axis, and other symbols are explained in the text.}
\label{figS_model}
\end{figure}

In the ``Parallel to Rotation'' model, we relate points at the ice-ocean interface to those on the sea floor that lie on the same line parallel to the axis of rotation (Figure~\ref{figS_model}(a)):

\begin{equation} \label{eq:Heat3}
    r_t \cos{\theta_t} = r_b \cos{\theta_b},
\end{equation}

\noindent where the subscript $t$ indicates the top of the ocean and the subscript $b$ indicates the bottom of the ocean, $r$ is radius and $\theta$ is latitude. Setting $\theta_b=0$ we can solve for the latitude of the tangent cylinder at the top of the ocean:

\begin{equation} \label{eq:theta_c}
    \left| \theta_t^{c_1} \right| = \cos^{-1} ({r_b / r_t}).
\end{equation}

If the bottom heating is entirely transported to the surface by the columnized convection, we moreover find

\begin{equation} \label{eq:Heat4}
    Q_b(\theta_b)r_b^2 \cos{\theta_b} d\theta_b d\lambda =  Q_t(\theta_t)r_t^2 \cos{\theta_t} d\theta_t d\lambda
\end{equation}

\noindent where $Q$ is the heat flux. From Equation~(\ref{eq:Heat3}) we get $d\theta_b/d\theta_t =  (r_t \sin{\theta_t})/(r_b\sin{\theta_b})$, so that Equation~(\ref{eq:Heat4}) can be solved for the heat flux at the ocean surface, $Q_t(\theta_t)$, inside the tangent cylinder. The surface heat flux is zero outside the tangent cylinder, so that it can generally be written as

\begin{equation} \label{eq:Heat5}
Q_t(\theta_t) = \left\{
\begin{array}{ccl}
0, & & {\mbox{if }\left| \theta_t \right|<\left| \theta_t^{c_1} \right|,}\\
Q_b(\theta_b)\frac{\sin{\theta_t}}{\sin{\theta_b}}, & & {\mbox{if }\left| \theta_t \right|>\left| \theta_t^{c_1} \right|,}
\end{array} \right.
\end{equation}

\noindent where $\theta_b$ is determined by Equation~(\ref{eq:Heat3}) and $\theta_t^{c_1}$ is given by Equation~(\ref{eq:theta_c}). Notice that Equation~(\ref{eq:Heat5}) has a singularity at the latitude where the surface intersects with the tangent cylinder.

In the ``Perpendicular to Rotation'' model (Figure~\ref{figS_model}(b)), following a similar argument as in the ``Parallel to Rotation'' model, we have

\begin{equation} \label{eq:Heat32}
    r_t \sin{\theta_t} = r_b \sin{\theta_b},
\end{equation}

\noindent which gives a critical latitude

\begin{equation} \label{eq:theta_c2}
    \left| \theta_t^{c_2} \right| = \sin^{-1} ({r_b / r_t}).
\end{equation}

By substituting Equation~(\ref{eq:Heat32}) into Equation~(\ref{eq:Heat4}) (which is also valid in this model), we have

\begin{equation} \label{eq:Heat52}
Q_t(\theta_t) = \left\{
\begin{array}{ccl}
Q_b(\theta_b)\frac{r_b}{r_t}, & & {\mbox{if }\left| \theta_t \right|<\left| \theta_t^{c_2} \right|,}\\
0, & & {\mbox{if }\left| \theta_t \right|>\left| \theta_t^{c_2} \right|,}
\end{array} \right.
\end{equation}

\noindent where $\theta_b$ is determined by Equation~(\ref{eq:Heat32}) and $\theta_t^{c_2}$ is given by Equation~(\ref{eq:theta_c2}).

In the ``Parallel to Gravity'' model, heat fluxes at the sea floor can be matched directly to surface fluxes at the same latitude, reduced only by a factor that accounts for the radial increase in surface area:

\begin{equation} \label{eq:Heat2}
   Q_t(\theta) = Q_b(\theta)\frac{r_b^2}{r_t^2}.
\end{equation}

In the ``Perpendicular to Gravity'' model, the surface heat flux is independent of latitude and simply given by the total heat flux divided by the surface area:

\begin{equation} \label{eq:Heat6}
    Q_t(\theta) = \frac{2\pi \int_{-\pi/2}^{\pi/2}Q_b(\theta')r_b^2 \cos{\theta'} d\theta'}{4\pi r_t^2}.
\end{equation}

Note that the detailed patterns of the surface heat flux vary with the bottom heat flux patterns, except for the uniform surface flux in the model where mixing is ``Perpendicular to Gravity'', which is independent of the bottom heat flux pattern. However, Equations (\ref{eq:Heat5}), (\ref{eq:Heat52}), (\ref{eq:Heat2}) and (\ref{eq:Heat6}) can be used to compute the surface heat fluxes in the respective limit cases for arbitrary bottom heat flux distributions, and many of the main features of the surface heat flux distribution are robust for moderate changes in the bottom heat flux pattern.

\section{Additional Simulation Details} \label{si:model}

\setcounter{equation}{0}

Due to the deep ocean (40~km) compared to the planetary radius (252~km), our GCM simulations do not use the thin-shell approximation typically applied in Earth-like simulations \citep{adcroft2018mitgcm}. As a result, the variation of grid cell area with depths at the same latitude and longitude is taken into consideration, which allows vertical fluxes to be calculated more precisely. The vertical variation of gravity is also taken into account. We calculate the gravity as a function of depth as

\begin{equation} \label{eqs:g}
    g(r) = g_c \frac{r_c^2}{r^2} + \frac{4\pi G\rho_w}{r^2} \int_{r_c}^{r} r'^{2}dr',
\end{equation}

\noindent where $G$ is the gravitational constant, $r_c$ is the radius at the surface of the solid core, i.e. the bottom of the ocean, $\rho_w$=1000~kg~m$^{-3}$ is the density of water (and also ice, for simplicity) and $g_c = 0.127$~m~s$^{-2}$ is the gravity at the bottom of the ocean, chosen such as to keep the gravity at the surface of the ice shell $g(r_s)$ to be the observed value of 0.113~m~s$^{-2}$. The vertical profile of the gravity is shown in Figure~\ref{figS_gravity}(a). 

\begin{figure}[b]
\centering
\includegraphics[width=0.8\linewidth]{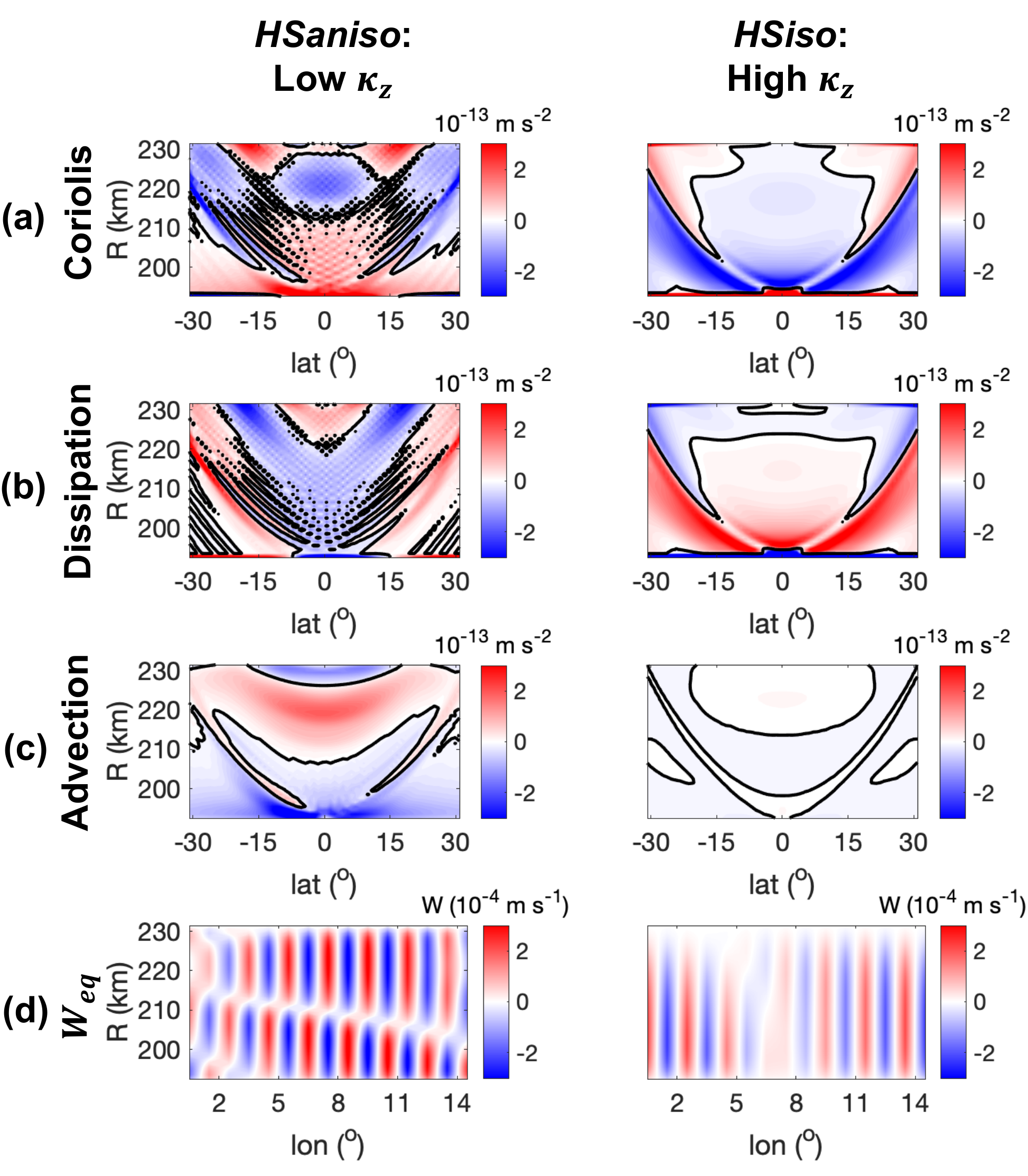}
\caption{Momentum budget and equatorial eddies in \textit{HSaniso} (left) and \textit{HSiso} (right). Panels (a)-(c) show ten-year-averaged zonal-mean zonal acceleration due to Coriolis force, dissipation, and advection terms, respectively. Black lines in (a)-(c) are zero contour lines. Panel (d) shows snapshots of vertical velocity in the equatorial plane. Note that the equatorial convective patterns are grid-scale because the horizontal resolution is not able to resolve the intrinsic scale of the convective plumes, but within the constraint imposed by this limitation, the equatorial velocity structure in \textit{HSaniso} shows a vertical tilt, which is associated with upward momentum transport.}
\label{figS_superrotation}
\end{figure}

Various studies have estimated the amplitude and patterns of tidal dissipation in the solid core on Enceladus \citep[e.g.][]{beuthe2013spatial, choblet2017powering, beuthe2018enceladus, beuthe2019enceladus}. Since small variations in the bottom heat flux pattern are not expected to affect the main results of this study, we assume a simplified bottom heating pattern that reflects the tidal disturbing potential. The global total energy input is assumed to be 20~GW, with the maximum heat flux twice of the minimum \citep{choblet2017powering}. The magnitude of the bottom heat flux may affect the convective mixing time scale ($\tau_{conv} \propto Q^{-1/2}$; Equation~(\ref{eq:convection})), the depth of the stratified layer ($H \propto Q^{-1}$; (Equation~(\ref{eq:H})), and the diffusive time scale in the stratified layer ($\tau_{diff} \propto Q^{-2}$; (Equation~(\ref{eq:diffusion})). Generally, a larger bottom heat flux will result in a shorter vertical convective mixing time scale, a shallower stratified layer and hence a shorter vertical diffusive mixing time scale in that layer. However, unless the bottom heat flux differs by orders of magnitude from the value assumed here, our main conclusions are expected to be robust. The heating pattern is then suggested to be:

\begin{equation} \label{eq:Q1}
    Q_{core}(\Theta,\lambda)=\frac{F_{total}[1/2 \, Y_{20}(\Theta)-1/4 \, Y_{22}(\Theta,\lambda)+C_0]}{4\pi C_0 r_c^2},
\end{equation}

\noindent where $F_{total}$= 20~GW is the total tidal heating, $C_0=(0.5+\sqrt{3}/8) \, \sqrt{5/\pi} \approx 0.904$ is a constant, $\Theta$ is co-latitude, $\lambda$ is longitude, and $Y_{20}$ and $Y_{22}$ are degree-2 spherical harmonic functions. Since we are only simulating part of the zonal range of the global ocean (15$^\circ$ in longitude), we cannot adequately simulate zonal variations. We therefore apply a zonally symmetric heating profile given by the zonal mean of Equation (\ref{eq:Q1}), which is (Figure~\ref{figS_gravity}(b)):

\begin{equation} \label{eq:Q}
    Q_{core}(\Theta)=\frac{F_{total}[1/2 \, Y_{20}(\Theta)+C_0]}{4\pi C_0 r_c^2}.
\end{equation}

In simulations with varying horizontal resolution, the viscosity coefficient is modified such that the Kolmogorov scale is proportional to the grid scale \citep{vallis2017atmospheric}. The Kolmogorov scale $L_\nu$ is the length scale at which the viscosity becomes important, and scales as:

\begin{equation}
    L_\nu \sim \varepsilon^{-1/4}\nu^{3/4},
\end{equation}

\noindent where $\varepsilon$ is the turbulent energy cascade rate, which should be independent of resolution, and $\nu$ is the turbulent viscosity. Setting $L_\nu \sim L_{grid}$, we can get the relationship:

\begin{equation}
    \nu \propto L_{grid}^{4/3}.
\end{equation}

\noindent We here choose $L_{grid}$ as the horizontal grid scale, which in our model is generally larger than the vertical grid scale. The horizontal grid length in case \textit{HSiso05} is half of that in case \textit{HSiso}, so that the viscosity is correspondingly changed from 2.5~m$^{2}$~s$^{-1}$ to $2.5\times 0.5^{4/3}\approx 1.0$~m$^{2}$~s$^{-1}$. We set other parameters in case \textit{HSiso05} by keeping the diffusivity and viscosity isotropic and $Pr$=10 (Table \ref{tb:experiments}).

\section{Mechanisms for Superrotation} \label{si:superrotation}

\setcounter{equation}{0}

There exists superrotation in the upper ocean in \textit{HSaniso} but not in \textit{HSiso} and \textit{HSiso05}. Here we compare \textit{HSaniso} and \textit{HSiso} to analyze the mechanism for equatorial superrotation. The two isotropic simulations, \textit{HSiso} and \textit{HSiso05}, are similar (not shown).

The zonal-mean zonal momentum budget is \citep{vallis2017atmospheric}:

\begin{eqnarray}\label{eq:zonal}
\frac{\partial \overline{u}}{\partial t} &=& 2\Omega(\overline{v}\sin{\theta} -\overline{w}\cos{\theta}) \nonumber \\
&+& \overline{F_x} \nonumber \\
&-& \overline{\frac{v}{a \cos{\theta}} \frac{\partial}{\partial \theta} (u\cos{\theta})} - \overline{{w} \frac{\partial u}{\partial r}}-\overline{\frac{uw}{r}} + \overline{\frac{uv \tan{\theta}}{r}},
\end{eqnarray}

\noindent where the overbar indicates the zonal and temporal average; $u$, $v$, $w$ are zonal, meridional and vertical velocity, respectively; $\theta$ is latitude, and $F_x$ is dissipation. On the right-hand-side, the terms in the first line are the Coriolis force terms, the term in the second line is the dissipation term, and the terms in the third line are the advection terms (including metric terms). In an equilibrium state, the left-hand-side should be zero.

Superrotation in \textit{HSaniso}, which is largest around $r = 219$~km (around 13~km in depth), is driven by momentum flux convergence (Figure~\ref{figS_superrotation}(c)) and counteracted by Coriolis terms and frictional dissipation (Figure~\ref{figS_superrotation}(a) \& (b)). In \textit{HSiso}, however, such a momentum flux convergence is absent. A cross section of vertical velocity along the equator (Figure~\ref{figS_superrotation}(d)) shows that the upward momentum flux in \textit{HSaniso} is driven by vertically tilted eddies \citep{aurnou2001strong, kaspi2008turbulent}. Although similar eddies appear to exist in \textit{HSiso}, they are weaker and not tilted, and hence do not carry momentum upwards.

\bibliography{Enceladus_Ocean}{}
\bibliographystyle{aasjournal}

\end{document}